\documentclass[aps,prx,reprint,superscriptaddress,floatfix]{revtex4-2}

\usepackage{amssymb}
\usepackage{graphicx}   
\usepackage{color}      
\usepackage{hyperref}   

\def\lsco{La$_{2-x}$Sr$_x$CuO$_4$}

\def\lnsco{La$_{1.6-x}$Nd$_{0.4}$Sr$_x$CuO$_4$}

\def\ybco{YBa$_2$Cu$_3$O$_{6+x}$}

\def\bscco{Bi$_2$Sr$_2$CaCu$_2$O$_{8+\delta}$}

\def\newr{\color{black}}

\begin{document}

\title{Strongly-overdoped \lsco: Evidence for Josephson-coupled grains of strongly-correlated superconductor}

\author{Yangmu Li}
\altaffiliation{Current address: Beijing National Laboratory for Condensed Matter Physics, Institute of Physics, Chinese Academy of Sciences, Beijing 100190, People's Republic of China.}
\affiliation{Condensed Matter Physics and Materials Science Division, Brookhaven National Laboratory, Upton, New York 11973-5000, USA}
\author{A. Sapkota}
\altaffiliation{Current address: Ames Laboratory, Iowa State University, Ames, IA 50011, USA.}
\affiliation{Condensed Matter Physics and Materials Science Division, Brookhaven National Laboratory, Upton, New York 11973-5000, USA}
\author{P. M. Lozano}
\affiliation{Condensed Matter Physics and Materials Science Division, Brookhaven National Laboratory, Upton, New York 11973-5000, USA}
\affiliation{Department of Physics and Astronomy, Stony Brook University, Stony Brook, NY 11794-3800, USA}
\author{Zengyi Du}
\affiliation{Condensed Matter Physics and Materials Science Division, Brookhaven National Laboratory, Upton, New York 11973-5000, USA}
\author{Hui Li}
\affiliation{Condensed Matter Physics and Materials Science Division, Brookhaven National Laboratory, Upton, New York 11973-5000, USA}
\affiliation{Department of Physics and Astronomy, Stony Brook University, Stony Brook, NY 11794-3800, USA}
\author{Zebin Wu}
\author{Asish K. Kundu}
\author{\newr{R. J. Koch}}
\author{\newr{Lijun Wu}}
\affiliation{Condensed Matter Physics and Materials Science Division, Brookhaven National Laboratory, Upton, New York 11973-5000, USA}
\author{B. L. Winn}
\author{Songxue Chi}
\author{M. Matsuda}
\author{M. Frontzek}
\affiliation{Neutron Scattering Division, Oak Ridge National Laboratory, Oak Ridge, Tennessee 37831, USA}
\author{{\newr E. S. Bo\v{z}in}}
\author{{\newr Yimei Zhu}}
\author{I. Bo\v{z}ovi\'c}
\affiliation{Condensed Matter Physics and Materials Science Division, Brookhaven National Laboratory, Upton, New York 11973-5000, USA}
\author{Abhay N. Pasupathy}
\affiliation{Condensed Matter Physics and Materials Science Division, Brookhaven National Laboratory, Upton, New York 11973-5000, USA}
\affiliation{Department of Physics, Columbia University, New York, NY 10027, USA}
\author{Ilya K. Drozdov}
\author{Kazuhiro Fujita}
\author{G. D. Gu}
\author{I. A. Zaliznyak}
\affiliation{Condensed Matter Physics and Materials Science Division, Brookhaven National Laboratory, Upton, New York 11973-5000, USA}
\author{Qiang Li}
\affiliation{Condensed Matter Physics and Materials Science Division, Brookhaven National Laboratory, Upton, New York 11973-5000, USA}
\affiliation{Department of Physics and Astronomy, Stony Brook University, Stony Brook, NY 11794-3800, USA}
\author{J. M. Tranquada}
\email{jtran@bnl.gov}
\affiliation{Condensed Matter Physics and Materials Science Division, Brookhaven National Laboratory, Upton, New York 11973-5000, USA}

\date{\today} 

\begin{abstract}
The interpretation of how superconductivity disappears in cuprates at large hole doping has been controversial.  To address this issue, we present an experimental study of single-crystal and thin film samples of \lsco\ (LSCO) with $x\ge0.25$.  In particular, measurements of bulk susceptibility on LSCO crystals with $x=0.25$ indicate an onset of diamagnetism at $T_{c1}=38.5$~K, with a sharp transition to a phase with full bulk shielding at $T_{c2}=18$~K, independent of field direction.  Strikingly, the in-plane resistivity only goes to zero at $T_{c2}$.  Inelastic neutron scattering on $x=0.25$ crystals confirms the presence of low-energy incommensurate magnetic excitations with reduced strength compared to lower doping levels.  The ratio of the spin gap to $T_{c2}$ is anomalously large.  Our results are consistent with a theoretical prediction for strongly overdoped cuprates by Spivak, Oreto, and Kivelson, in which superconductivity initially develops within disconnected self-organized grains characterized by a reduced hole concentration, with bulk superconductivity occurring only after superconductivity is induced by proximity effect in the surrounding medium of higher hole concentration.  Beyond the superconducting-to-metal transition, local differential conductance measurements on an LSCO thin film suggest that regions with pairing correlations survive, but are too dilute to support superconducting order.  Future experiments will be needed to test the degree to which these results apply to overdoped cuprates in general.
\end{abstract}

\maketitle

\section{Introduction}

While there is a common belief that magnetic correlations play a role in the pairing mechanism of cuprate superconductors \cite{keim15,scal12a}, the details remain controversial.  To make progress, it is helpful to test for experimental signatures that can distinguish between different theoretical predictions.   For example, early studies of the doping dependence of the Hall effect \cite{taka89b} and optical conductivity \cite{uchi91} indicated that the carrier concentration at low temperatures initially grows in proportion to the doped-hole density, $p$,
consistent with the effective localization of the $3d_{x^2-y^2}$ hole on each Cu site due to strong onsite Coulomb repulsion and with the surviving antiferromagnetic spin correlations \cite{ande87,emer87,kast98,birg06}.  The presence of strong interactions leads to the question of why $T_c$ is relatively low \cite{lee87}.  Emery and Kivelson \cite{emer95a} argued persuasively that the $T_c$ of underdoped cuprates is constrained by the small superconducting phase stiffness, which is limited by the low carrier density and consequent low superfluid density (inferred from muon spin rotation studies \cite{uemu91}), rather than by the onset of pairing correlations; they assumed that the decrease of $T_c$ with overdoping would be controlled by the mean-field behavior predicted by BCS-Eliashberg theory \cite{schr64}.  This latter assumption has now become a matter of controversy, given experimental observations that the carrier concentration crosses over from $p$ to $1+p$ with overdoping \cite{bado16,putz21} while $T_c$ and the superfluid density drop toward zero \cite{bozo16}.    While disorder seems to be relevant to understanding the superconductor-to-metal transition at $p=p_c\sim0.3$ \cite{ayre21}, there is no agreement on the role or nature of the disorder.

Early studies of Meissner diamagnetism (measured on cooling in a small magnetic field) in polycrystalline samples of overdoped \lsco\ (LSCO) found evidence for an onset above 35 K but a bulk-like transition at a much lower temperature \cite{torr88,taka92a}; for $p=x=0.25$, the bulk transition was below 20~K \cite{taka92a}.  Similarly, $c$-axis-polarized optical reflectivity measurements on LSCO single crystals observed very little superconducting response for $x=0.25$, leading to the suggestion that the superconductivity detected by resistivity and magnetic susceptibility measurements on the same sample (but never published) was not a bulk response \cite{uchi96} (but see \cite{naga93} for a different opinion).  Later studies of magnetic susceptibility and specific heat on overdoped LSCO crystals led to proposals of microscopic phase separation between superconducting and metallic phases \cite{wen00,tana05,tana07,wang07}.  Cu nuclear magnetic resonance (NMR) measurements indicated an increase in the density of unpaired charge carriers for $x\geq0.20$ \cite{ohsu94}.

A different perspective on overdoped LSCO came with the study of thin films.  Measurements by the two-coil mutual inductance technique indicated very sharp superconducting transitions, even as the superconducting transition temperature, $T_c$, and the superfluid density, $n_s$, decreased with growing doped-hole density, $p$ \cite{lemb10,lemb11}.  A definitive study of a very large number of high-quality LSCO thin films confirmed the sharp transitions for overdoped samples \cite{bozo16}.   Tests on representative films with a scanning probe indicate uniformity of the superconductivity down to a length scale of microns \cite{herr21}.  Terahertz spectroscopy measurements on the same superconducting samples provide direct evidence of a significant fraction of uncondensed carriers that grows with $p$ \cite{mahm19}.  While the single sharp transition for each $p$ appears different from the results on overdoped bulk samples, the THz spectroscopy on films \cite{mahm19} and the specific heat results on sintered powders \cite{momo94} and crystals \cite{wang07} both show the presence of free carriers coexisting with superconductivity.

The main results of \cite{bozo16} are: 1) the temperature-dependence of $n_s$ is consistent with expectations for a clean $d$-wave superconductor \cite{hard93} at each doping, and 2) both $T_c$ and $n_s$ decrease with growing $p$, indicating a decreasing fraction of charge carriers participating in pairing.   Given that the normal-state charge transport appears to be relatively conventional in the very overdoped metallic regime \cite{coop09}, some theorists find it tempting to apply weak-coupling BCS theory for $d$-wave superconductivity, with the assumption of well-defined quasiparticles in the normal state.  The coexistence of paired and unpaired electrons can be reproduced with a mean-field treatment of disorder attributed to impurity scattering, and this approach has been used to model the doping dependence of some experimental quantities \cite{leeh17,leeh18,leeh20,khod19,wang22a}; note that the assumption of an important role for disorder is in conflict with the conclusions of \cite{bozo16}.  One challenge is that, for $d$-wave pairing, the disorder modifies the temperature dependence of the superfluid density, making it difficult to describe the experimental results \cite{bozo18,mahm22}.  Another challenge is that studies of the doping dependence of the normal-state resistivity in LSCO across the overdoped regime are not consistent with an assumption of conventional transport plus a scattering rate that increases with $p$ \cite{naka03,huss08,coop09,huss18,bozo18}.  

That last observation raises the issue of the strength of interactions in superconducting cuprates and the relevance of weak-coupling approaches.   In the underdoped regime, the cuprates clearly behave as doped antiferromagnets \cite{keim15}, and the competition between the kinetic energy of doped holes and superexchange energy between Cu spins is largely responsible for the unusual behavior of the normal-state electronic response \cite{frad15,tran21a}, such as the observation that only the dopant-induced holes act as charge carriers \cite{ando04}.  Neutron scattering studies on cuprates, and especially LSCO, have demonstrated that the spectral weight of antiferromagnetic (AF) spin fluctuations decreases with doping \cite{kive03,waki04,birg06,waki07b,ma21}.  In a consistent fashion, the strong-scattering component (``mid-IR'') of optical conductivity has been observed to decrease and disappear as $T_c$ goes to zero with over doping \cite{mich21}; similarly, the decrease in the strong-correlation effects is reflected in the observations that the carrier density measured by the Hall effect crosses over from $p$ towards $1+p$ \cite{bala09,bado16,putz21}.  From the theoretical side, a recent calculation based on the Hubbard model, with onsite repulsion $U$, nearest-neighbor hopping $t$, and $U/t=7$, explored the overdoped regime, finding that both the averaged pairing strength and the strength of the relevant antiferromagnetic correlations decrease toward zero as $p$ grows \cite{maie20}.   

The crossover to the strongly-overdoped regime occurs at $p^\ast\sim0.19$, which corresponds to the pseudogap critical point \cite{tall01,keim15}.  The pseudogap corresponds to the depression of low-energy electronic states in the normal state at wave vectors where the superconducting $d$-wave gap is largest (``antinodal'' states) \cite{vish12}; these are also the states most impacted by the presence of AF spin correlations \cite{wu22}. The pseudogap is a characteristic of the normal state for $p<p^\ast$ and develops below a temperature $T^\ast$.  One measure of $T^\ast$ is the peak in the bulk magnetic susceptibility, $\chi_s$; the decrease of $\chi_s$ with cooling is consistent with the development of short-range antiferromagnetic spin correlations \cite{john89,huck08,maie19}.  In LSCO, the temperature of the peak in $\chi_s(T)$ decreases with increasing $p$, and a Curie-like contribution appears for $p>p^\ast$ \cite{naka94}.  Now, neutron scattering measurements on LSCO show that the AF spin excitations are gapped for $x>0.13$ \cite{chan08}, although the gap can be closed with a sufficiently strong magnetic field.  A recent high-field study detecting glassy AF correlations by NMR and ultrasound found that the field-induced glassy order disappears at $p\sim p^\ast$ \cite{frac20}.  These observations suggest that $p^\ast$ is something like a percolation limit for AF correlations, and that the AF spin correlations no longer extend throughout the CuO$_2$ planes for $p>p^\ast$, but instead become isolated.

\begin{figure}[t]
 \centering
    \includegraphics[width=0.95\columnwidth]{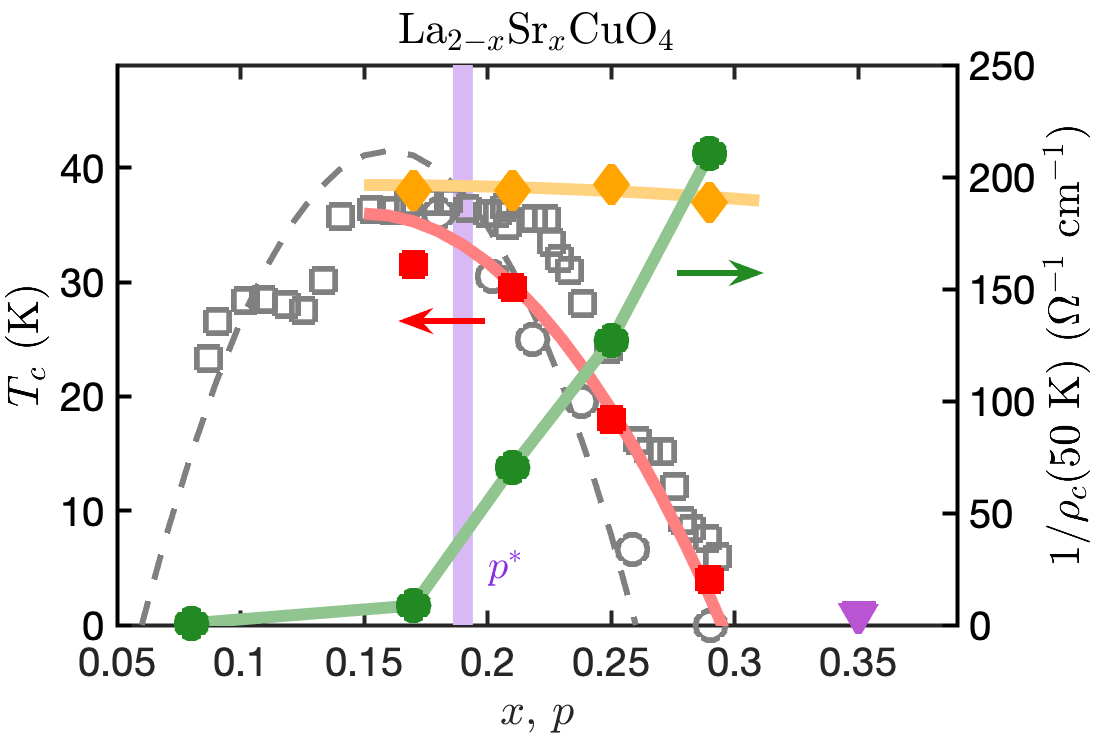}
    \caption{\label{fg:tc}  Superconducting transition temperatures as a function of the doped-hole concentration in \lsco\ samples.  The orange diamonds indicate onset of the Meissner effect in the single crystals, while the red squares indicate where the in-plane resistivity goes to zero; lines through the symbols are guides to the eye.  The violet triangle indicates the estimated $p$ for the metallic LSCO thin film, which had no bulk superconducting transition above 4.2~K.  The gray line is the behavior of $T_c$ vs.\ $p$ assumed in \cite{bozo16}, with a form similar to that proposed in \cite{tall95}.  The gray circles {\newr (squares)} correspond to single crystals studied in \cite{wang07} {\newr (\cite{ikeu03}).  Green filled circles represent $1/\rho_c$ at 50~K, the inverse of the $c$-axis resistivity, a measure of the $c$-axis conductivity in the normal state. } }
\end{figure}

{\newr
As an illustration of the change in electronic character across $p^\ast$, the green circles in Fig.~\ref{fg:tc} indicate a measure of the normal-state conductivity perpendicular to the CuO$_2$ planes.  LSCO is essentially insulating along the $c$ axis for $p<p^\ast$ \cite{naka93b,uchi96}, where AF correlations are dominant.  Electronic dispersion along $c$ only becomes detectable for $p>p^\ast$ \cite{hori18,zhon22}, as the chemical potential moves into the large density of antinodal states.  While the specific heat shows a strong peak at $p\sim p^\ast$ \cite{mich19}, low-energy incommensurate AF spin correlations are strong there \cite{zhu22} (though with no sign of quantum critical behavior \cite{li18}), and antinodal states at $p=0.23 > p^\ast$ retain a large self-energy \cite{chan13}.  Hence, the gradual growth of conductivity along the $c$ axis for $p>p^\ast$ appears to reflect the gradual reduction of strong-correlation effects.
}

Muon spin relaxation ($\mu$SR) studies on LSCO have found evidence for a more heterogeneous response to an applied magnetic field in the overdoped regime \cite{macd10}, with evidence for two distinct local environments \cite{kais12}.  An analysis of x-ray absorption spectra measured at the O $K$ edge in LSCO and Tl$_2$Ba$_2$CuO$_{6+\delta}$ finds evidence for reduced correlation effects for $p\gtrsim p^\ast$ \cite{peet09}.  A complementary picture has been derived from the analysis of magnetotransport measurements on several cuprate families \cite{ayre21}: the strongly dissipative behavior observed for $p<p^\ast$ crosses over to a combination of contributions characteristic of both conventional quasiparticles and strong dissipation, with the latter disappearing as $T_c \rightarrow 0$.   

One source of spatial inhomogeneity of the local charge density is the random distribution of the dopant ions.  In the case of LSCO, this involves ionic Sr$^{2+}$ randomly substituting for La$^{3+}$; the spatial inhomogeneity of the local doped-hole density has been inferred from measurements with local probes such as NMR \cite{sing02a,sing05} and scanning tunneling microscopy (STM) \cite{kato08}; inhomogeneity is also indicated by observations of nonlinear conductivity \cite{pelc18} and finite structural correlation lengths \cite{mosq09}.  Spivak, Oreto, and Kivelson \cite{spiv08} considered models of a composite of superconducting and metallic regions, where there is a quantum phase transition from superconductor to metal tuned by disorder.  For the case of $d$-wave superconductivity, there is a regime in which cooling leads initially to a metallic phase that contains dilute, locally-superconducting puddles, with a transition at lower temperature to a state having uniform superconducting order.  This picture, supported by recent numerical modeling \cite{li21c}, is distinct from phase separation; instead, stochastic effects lead to self-organized granular behavior.

In this paper, we report experimental measurements on {\newr several} overdoped LSCO samples,  superconducting single crystals with $x=0.25$ and 0.29 and a metallic thin film with $p\sim0.35$, and show  that the data are consistent with the scenario described above.  The superconducting transitions observed in these samples are summarized in Fig.~\ref{fg:tc}, together with results on related crystals to be presented separately \cite{li22p}.  At temperature $T_{c1}$ (diamonds), we observe the onset of diamagnetism in the crystal samples, while the in-plane resistivity goes to zero at the lower $T_{c2}$ (squares).  Note that the separation between these temperatures grows as $p$ exceeds $p^\ast$.  We will show for the crystal{\newr s} that the initial diamagnetism between $T_{c1}$ and $T_{c2}$ involves a small volume fraction, consistent with disconnected three-dimensional bubbles of superconductivity, while there is a sharp increase below $T_{c2}$ as the magnetic shielding grows rapidly to include the entire volume.  For the metallic, non-superconducting LSCO film doped beyond the superconductor-to-metal transition, spectroscopic-imaging STM (SI-STM) measurements provide evidence of residual, isolated regions with a response suggestive of pairing correlations at low temperature.

We also present measurements of the incommensurate antiferromagnetic spin excitations for the $x=0.25$ crystal.  From the change of the dynamic susceptibility with temperature, we identify a spin gap that, compared with $T_{c2}$, is much larger than what one finds near optimal doping \cite{li18}.  We argue that this is evidence that the local superconductivity is driven by correlated domains, and that the transition to bulk order requires the induction of superconductivity in the intervening less-correlated metallic regions by proximity effect \cite{spiv08}.

The rest of the paper is organized as follows.  In the following section, we describe the sample growth and characterizations, as well as the methods for measuring superconducting properties and the spin excitations.  The magnetic susceptibility and resistivity results are described in Sec.~III, followed by a presentation of the inelastic neutron scattering results in Sec.~IV.  SI-STM measurements are detailed in Sec.~V.  Further discussion of the results, their significance, and remaining questions appears in Sec.~VI.

\section{Materials and Methods}

\subsection{Sample growth and characterization}

\subsubsection{\newr Crystal growth}

The present LSCO $x=0.25$ crystal was grown by the same procedure used for two recently-studied $x=0.17$ and 0.21 compositions \cite{li18,miao21}.  A floating-zone furnace equipped with two ellipsoidal mirrors was used to grow the single crystal at a velocity of 1~mm/h under an air flow rate of 0.5 l/min. For each composition of  \lsco, the feed rod was prepared from powders of La$_2$O$_3$, SrCO$_3$ and CuO (99.99\%\ pure), combined in their appropriate metal ratios, together with 1\%\ extra CuO to compensate for loss due to evaporation during crystal growth.  The combined powders were treated by a repeated process of grinding in an agate mortar followed by calcination at 980~$^\circ$C (first round) and then 1050~$^\circ$C. The pressed feed rods were sintered for 72 h at 1300~$^\circ$C in air.  Each grown crystal rod was annealed in pure O$_2$ flow at 980~$^\circ$C for 200~h, followed by furnace cooling (which takes several hours).  The temperature-dependent crystal structure for the $x=0.17$, 0.21, and 0.25 crystals was determined by neutron diffraction, as described in App.~A.  A crystal with $x=0.29$ was grown and annealed in a similar fashion.

{\newr Although the low-temperature crystal structure  changes from orthorhombic to tetragonal with doping, we will use the notation of the orthorhombic cell for consistency, with $a\approx b \approx 5.32$~\AA, $c\approx 13.2$~\AA.  Momentum transfer {\bf Q} and reciprocal lattice vectors will be expressed in reciprocal lattice units (rlu), $(2\pi/a,2\pi/b,2\pi/c)$.  In these units, the AF wave vector corresponds to ${\bf Q}_{\rm AF}=(1,0,0)$ [or, equivalently, $(0,1,0)$].}

{\newr
\subsubsection{X-ray diffraction characterization}

\begin{figure}[b]
 \centering
    \includegraphics[width=0.95\columnwidth]{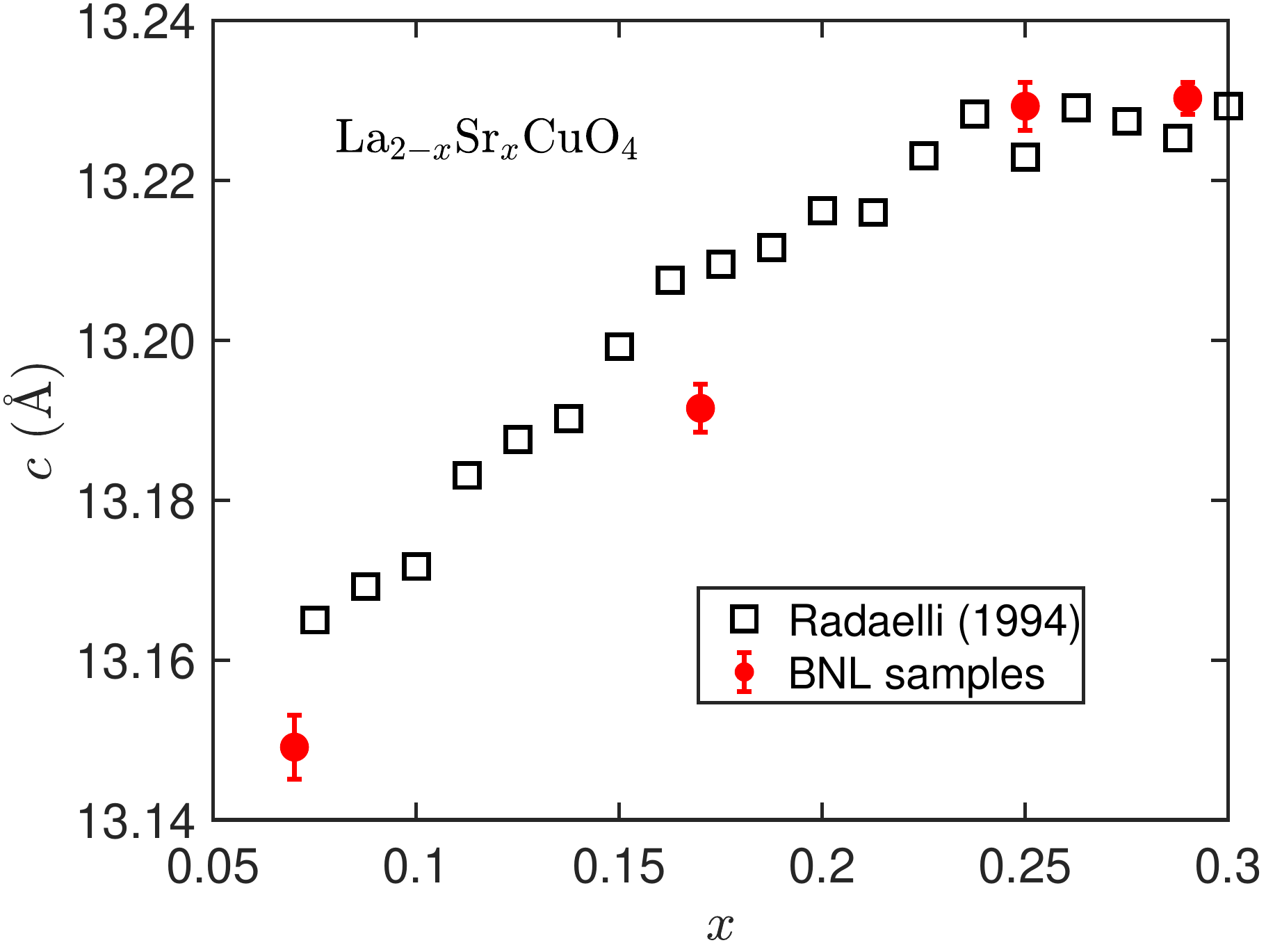}
    {\newr
    \caption{\label{fg:cvx}  Doping dependence of the $c$ lattice parameter of powdered pieces of single crystal, measured by high-resolution X-ray diffraction at 100 K (red filled circles), compared with powder neutron diffraction results at 70 K from Radaelli {\it et al.}\ \cite{rada94}.  The error bars on the x-ray results correspond to 10$\sigma$, thus enhanced so as to be visible. }
    }
\end{figure}

Small pieces of some of the single-crystal samples were ground to powder and measured by high-resolution X-ray powder diffraction at 11-BM, Advanced Photon Source, Argonne National Laboratory, with each sample cooled to 100~K.  The diffraction peaks are very narrow, but with ${\bf Q}$-dependent widths somewhat greater than those measured on a standard reference sample, which makes Rietveld refinement impractical.  Instead, the diffraction pattern was analyzed by the Le Bail method \cite{leba05}.  The results for the $c$ lattice parameter are plotted vs.\ nominal doping in Fig.~\ref{fg:cvx}, where they are compared with powder neutron diffraction results from Radaelli {\it et al.}\ \cite{rada94}; the samples in the latter study were oxygen annealed, measured to be very close to stoichiometric in oxygen for $x\lesssim0.3$, and single phase.   The value of $c$ grows roughly linearly with $x$ up to $x\sim0.2$, but gradually flattens out above that.  [The latter change correlates with evidence for a modification in the character of doped holes at large doping, including: 1) an observed saturation of the effective in-plane O $2p$ hole concentration measured by x-ray absorption spectroscopy \cite{peet09}; 2) indications for an enhanced role of Cu $3d_{z^2}$ orbitals from Compton scattering \cite{saku11}, angle-resolved photoemission \cite{matt18}, and recent calculations \cite{wata21}.]   Our $x=0.25$ and 0.29 samples showed some weak diffraction peaks consistent with La$_2$SrCu$_2$O$_6$ \cite{hink87,nguy80}; the volume fraction of the impurity phase is estimated to be $<3$\%. If we take the width of the (008) peak as providing an upper limit on the spread in composition, then for $x=0.25$ we have $\Delta Q \approx 4\times 10^{-5}$~\AA$^{-1}$, which gives $\Delta x < 0.001$ [where we take $\Delta x \approx  \Delta c/(0.2\,{\rm \AA})$ in the range $0.2\lesssim x\lesssim0.25$].  }

{\newr

\subsubsection{Transmission electron microscopy characterization}

\begin{figure*}[t]
 \centering
    \includegraphics[width=2\columnwidth]{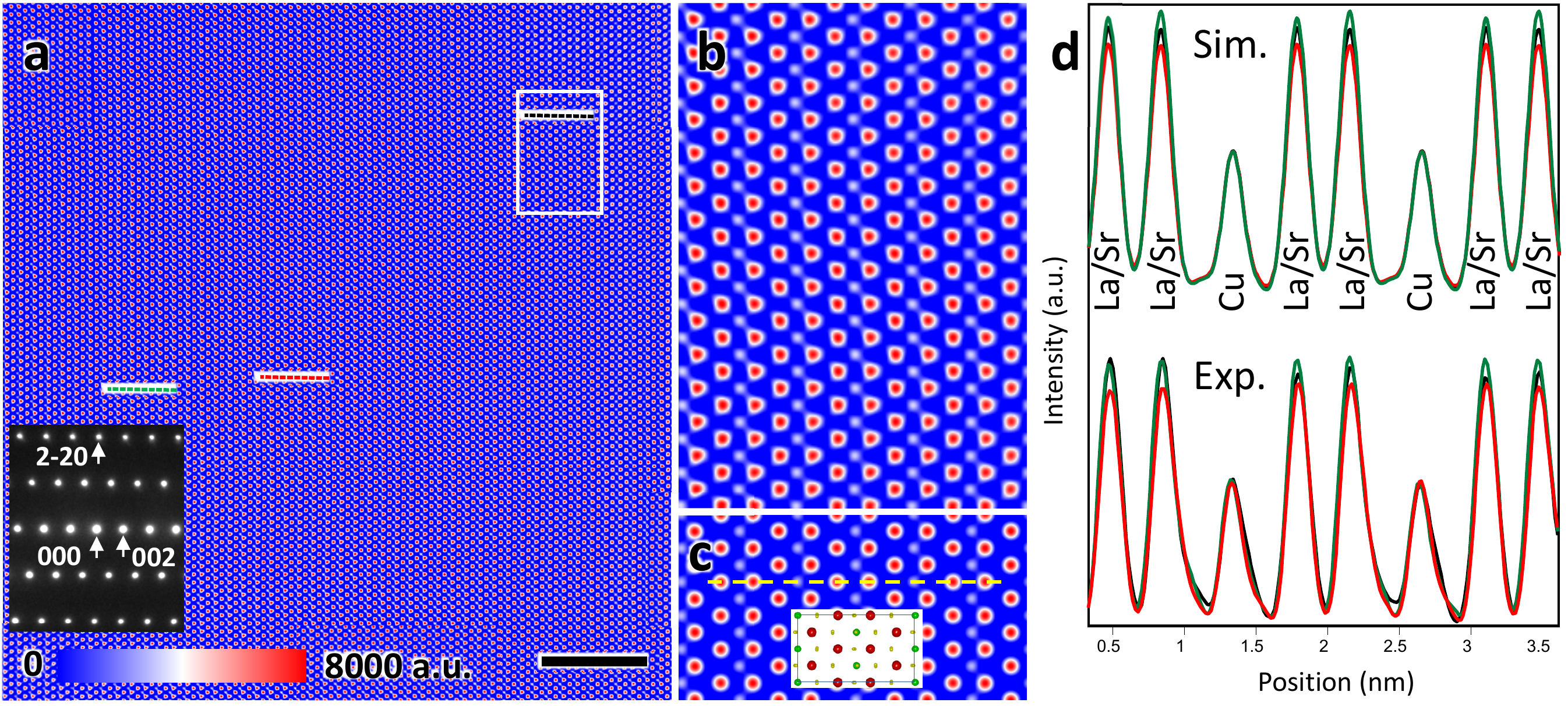}
    {\newr
    \caption{\label{fg:tem}  (a) STEM-HAADF image of LSCO $x=0.25$ crystal for an incident beam direction of [110]; scale bar\ $=5$~nm. The inset is the corresponding selected-area electron diffraction pattern from a large area (about 120 nm in diameter) with several reflections indexed.  (b) Magnified images from the area marked by the white rectangle in (a), showing the uniform intensity distribution at La/Sr and Cu sites. (c) Simulated STEM-HAADF image, as discussed in the text. The inset shows the [110] projection of the structure, with red, green and dark yellow spheres representing La/Sr, Cu and O, respectively.  (d) Intensity profiles from simulated images (top) with different compositions from yellow dashed scan line shown in (c) and experimental image (bottom) from the dashed lines shown in (a) with the corresponding colors.  The simulated intensity profiles are calculated for compositions of $x=0.18$ (green line), $x=0.25$ (black) and $x=0.32$ (red), respectively, all normalized to the Cu peak. }
    }
\end{figure*}

To further characterize the LSCO $x=0.25$ crystal, a piece was examined by scanning transmission electron microscopy (STEM) recorded by a high-angle annular dark field (HAADF) detector, as shown in Fig.~\ref{fg:tem}.   This technique allows atomic-scale resolution of columns of atoms and quantitative analysis of composition through simulation \cite{penn92,kirk98,wang22}.  The image has been filtered in frequency space by applying a periodic mask to suppress the background noise.  Both STEM-HAADF image (a) and selected-area electron diffraction pattern (inset) confirm that the sample is a single crystal, free from defects. In STEM-HAADF imaging, the contrast is approximately proportional to $Z^{1.7}$ along an atomic column, where $Z$ is the atomic number. The contrast of La/Sr columns is stronger than that of Cu columns, while the O is invisible.   An expanded view of a small region is shown in Fig.~\ref{fg:tem}(b), while a simulation is presented in (c).  The simulated image is based on the multi-slice method with frozen phonon approximation for the LSCO $x=0.25$ nominal composition with estimated thickness $t=25$ nm.  It has been convolved with a Gaussian function with full width at half maximum of 0.15 nm to take the probe shape into account.  Simulated (top) and experimental (bottom) intensity profiles are compared in Fig.~\ref{fg:tem}(d). 

\begin{figure}[b]
 \centering
    \includegraphics[width=0.95\columnwidth]{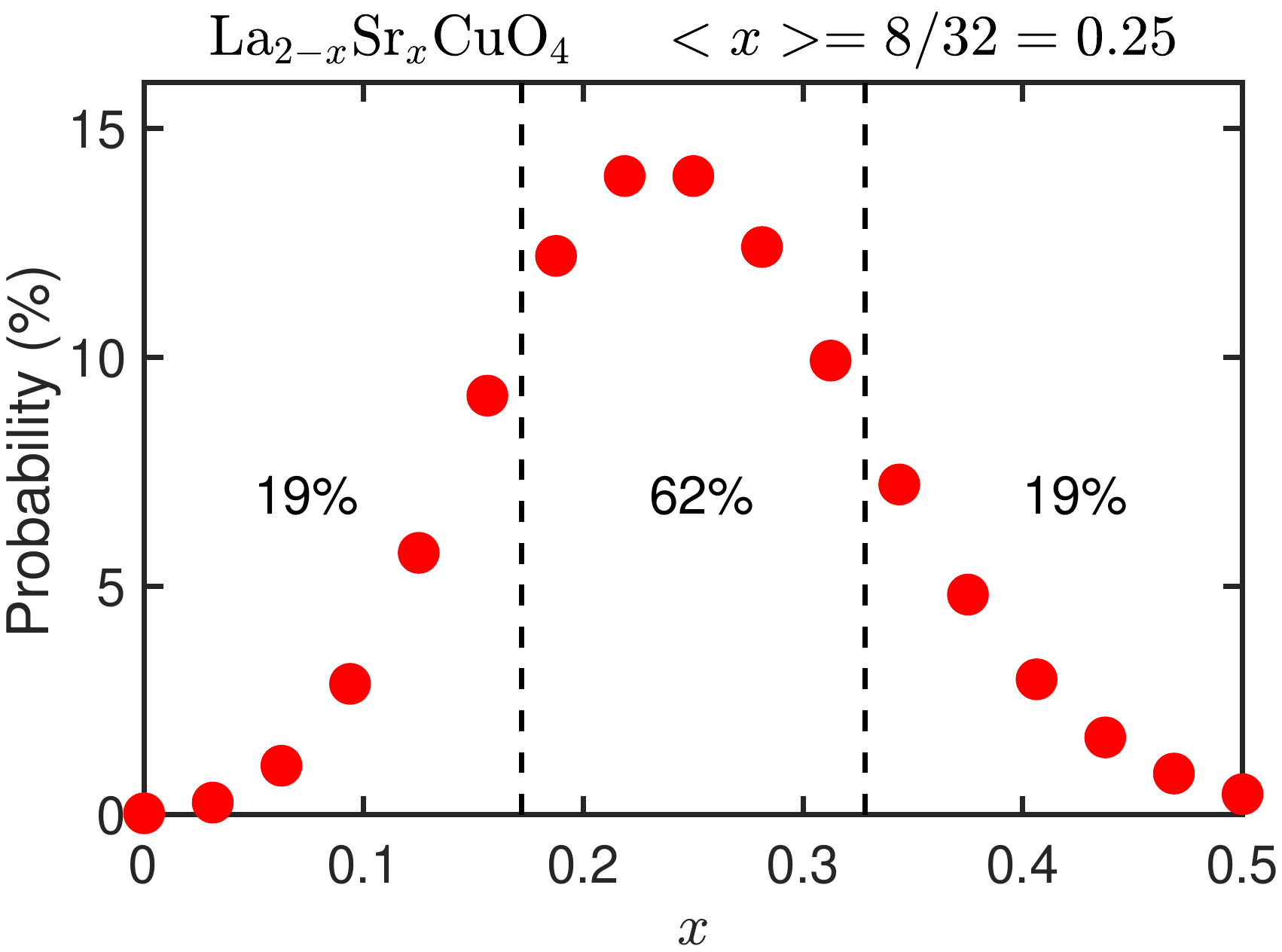}
    {\newr
    \caption{\label{fg:tprob}  Probability that an atomic column of 64 La sites will have a particular Sr content, assuming that the average is 8 Sr ions, corresponding to $x=0.25$. }
    }
\end{figure}

The peak intensities of La/Sr columns increase with the increase of La/Sr ratio.  The variations of the experimental peak intensities are consistent with simulations over the composition range $0.18\lesssim x\lesssim 0.32$, and compatible with an average of $x=0.25$.  While the spread of $x$ values may appear surprisingly large, it is actually perfectly consistent with a random distribution of Sr atoms on the La site with an average composition of $x=0.25$.  To see this, we start by noting that the sample thickness is approximately 64 unit cells, which means that each atomic column contains approximately 64 La sites.  For $x=0.25$, on average $1/8$ of the La sites are occupied by Sr, so that we expect an average of 8 Sr atoms per column.  The probability of finding columns with various compositions should follow the Poisson distribution, which we plot in Fig.~\ref{fg:tprob}.  There is a 62\%\ probability that the $x$ value observed in a given column will be within the range $0.18\lesssim x\lesssim 0.32$, consistent with the conclusion from the simulations.

To summarize, the STEM results indicate the absence of gross defects or elemental segregation, but demonstrate the relevance of stochastic effects.
}

\begin{figure*}[t]
 \centering
    \includegraphics[width=1.6\columnwidth]{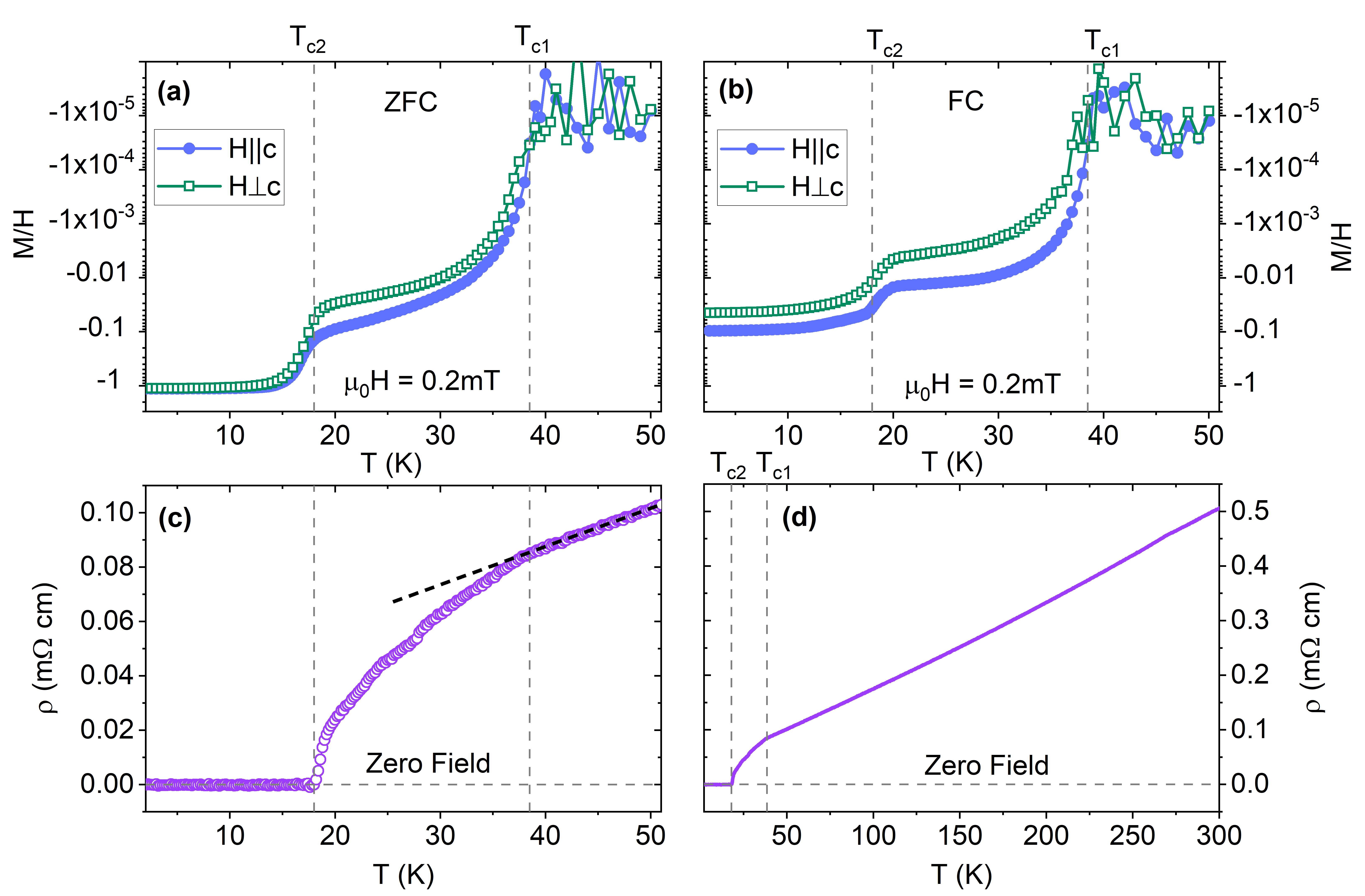}
    \caption{\label{fg:chi} Results for the LSCO $x=0.25$ crystal. (a) Volume susceptibility measured after zero-field cooling (ZFC) as a function of temperature in a magnetic field of 0.2~mT applied parallel to the $c$ axis (filled circles) or perpendicular to the $c$ axis (open squares).  (A constant value of $1\times10^{-5}$ has been subtracted from $M/H$ to accommodate plotting on a logarithmic scale, and magnitude has been corrected for the demagnetization factor.) (b) Same as (a) but for field-cooled (FC) measurements.
(c) In-plane resistivity in zero magnetic field; bold dashed line indicates normal-state trend.  (d)  In-plane resistivity over a larger temperature range.  Vertical dashed lines denote two characteristic temperatures: $T_{c1}=38.5$~K is the onset of diamagnetism and $T_{c2}=18.0$~K is the point at which the resistivity goes to zero (and bulk magnetic shielding onsets).}
\end{figure*}

\subsubsection{\newr Film growth}

To explore higher doping, we made use of an LSCO thin film grown by oxide molecular-beam epitaxy (OMBE) in the OASIS complex \cite{kim18}.  The film, with a thickness of 10 unit cells along the $c$ axis, was synthesized on a LaSrAlO$_4$ substrate by depositing La, Sr and Cu, which were sequentially shuttered to form the LSCO layer by layer in the presence of ozone. The ozone partial pressure was kept constant during the synthesis at $3\times 10^{-5}$~Torr, with the substrate temperature held at $T_s = 650~^\circ$C.  Mutual inductance measurements on the film later confirmed the absence of a superconducting transition down to 4.2~K.  

The nominal Sr concentration of the film is $x=0.4$, but a separate characterization is needed to determine the doped-hole concentration $p$, given uncertainty in the oxygen stoichiometry.  This was accomplished by transferring the sample {\it in vacuo} to the angle-resolved photoelectron spectroscopy (ARPES) chamber.  As discussed in App.~B, the ARPES measurements yielded an estimated hole concentration of $p\sim0.35$.  The sample was then transferred {\it in vacuo} to the STM chamber, where the spectroscopic imaging measurements were subsequently performed at $T=9$~K. 

{\newr A more complete STM study of LSCO films at several dopings is in progress and will be reported elsewhere \cite{LSCO_STM}.}

\subsection{Magnetization and transport}

To measure the temperature dependence of the magnetization of the LSCO $x=0.25$ sample, two small crystals were prepared, one each for field parallel and perpendicular to the $c$ axis.  Each crystal had the shape of a square plate, with sides of dimension 3 to 5 mm and plate thickness of 0.5 mm, oriented so that one of the long axes was parallel to the applied magnetic field.
The volume magnetic susceptibility (defined as $\chi= M/\mu_0 H$, where $M$ is the volume magnetization in Tesla, $\mu_0 H$ is the external magnetic field in Tesla) was measured in a Quantum Design Magnetic Properties Measurement System with a SQUID (superconducting quantum interference device) magnetometer.  The results, corrected for the demagnetization factor, are plotted in Fig.~\ref{fg:chi}(a) and (b). 

Transport measurements were carried out by the four-probe in-line method on a crystal oriented to determine the $ab$-plane resistivity, $\rho_{ab}$, in a 14 T Quantum Design Physical Properties Measurement System. The current contacts were made at the ends of each crystal along the long direction to ensure a uniform current flow throughout the entire sample; voltage contacts were made in direct contact with the $ab$-plane edges.  The results are presented in Fig.~\ref{fg:chi}(c) and (d).

The magnetization and resistivity results were reproduced on a second set of crystals.  Similar measurements have been performed on crystals cut from the LSCO $x=0.17$, 0.21, and 0.29 samples; a comparison of $x=0.29$ with 0.25 is {\newr presented in Sec.~\ref{sc:st}.  The inverse of $\rho_c$ at $T=50$~K for this series of crystals is shown in Fig.~\ref{fg:tc}, and more} complete data will be presented in detail elsewhere \cite{li22p}.

\subsection{Inelastic neutron scattering}

To characterize the antiferromagnetic spin correlations in the LSCO $x=0.25$ sample, inelastic neutron scattering measurements were performed on the HYSPEC instrument at the Spallation Neutron Source, Oak Ridge National Laboratory.  The sample consisted of 5 co-aligned crystals with a total mass of 32~g, oriented to have the $c$ axis vertical, perpendicular to the scattering plane.  It was mounted in a pumped-He cryostat, to allow a base temperature of 2~K.  All measurements were performed with an incident energy of 27~meV and a chopper frequency of 300 Hz.  To put the scattered intensity in absolute units, the data were calibrated to measurements on a hollow vanadium cylinder.

\section{Superconducting transitions}
\label{sc:st}

The magnetic susceptibility data in Fig.~\ref{fg:chi}(a) and (b) for LSCO $x=0.25$ are striking.  The development of superconducting order is characterized by two features: an initial onset of diamagnetism at $T_{c1}= 38.5$~K, associated with a response that gradually grows to a shielding volume of $\sim10$\%, followed by a bulk transition at $T_{c2} =18$~K that leads to full volume shielding.  In both regimes, the superconductivity is three-dimensional, with screening of magnetic fields observed for both orientations of the applied magnetic field.  The difference in strength of the diamagnetism with field orientation in the high-temperature phase correlates with the large anisotropy in magnetic penetration depth \cite{uchi96}.   

\begin{figure}[b]
 \centering
    \includegraphics[width=0.9\columnwidth]{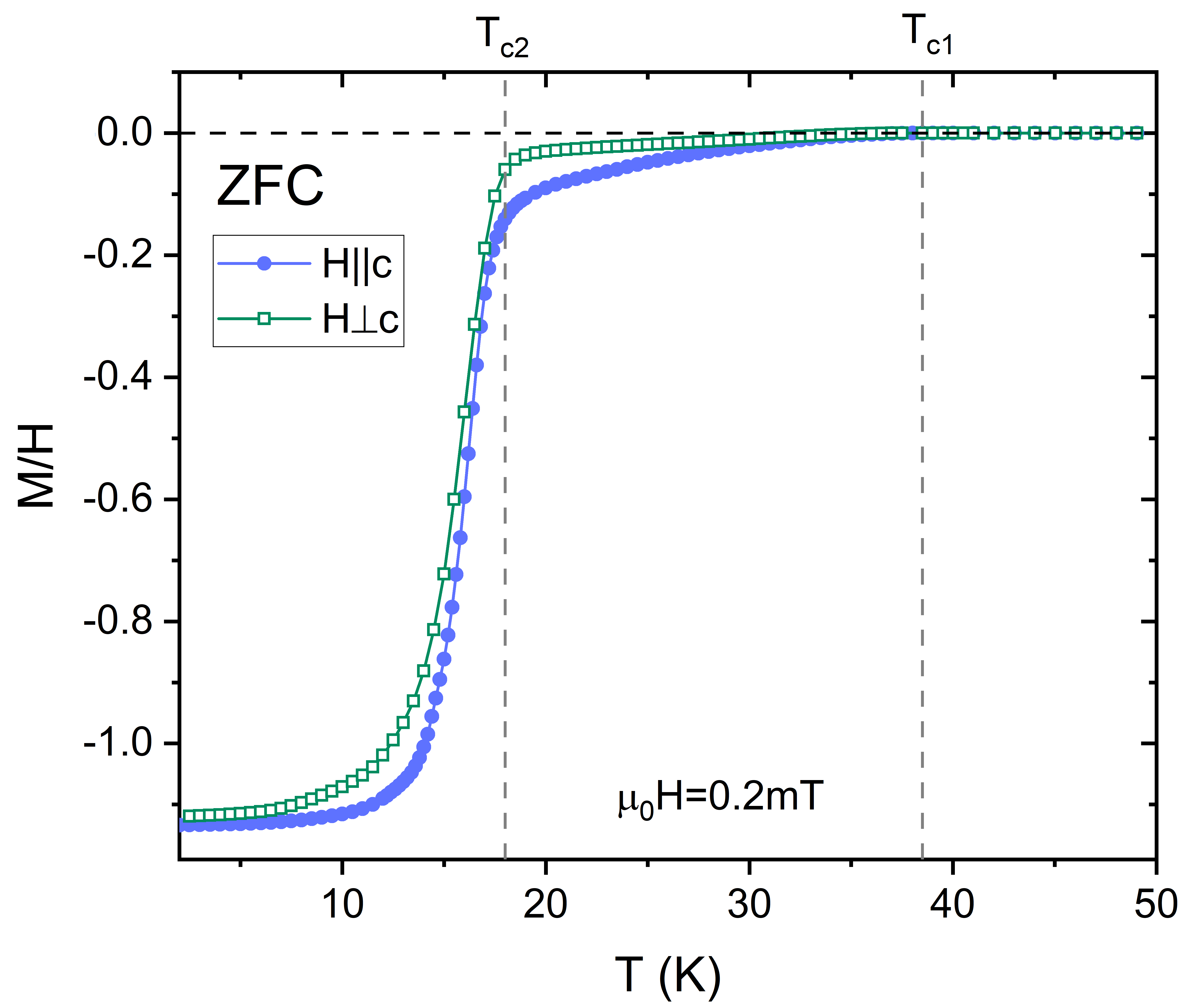}
    \caption{\label{fg:chilin}  ZFC volume susceptibility data for LSCO $x=0.25$ from Fig.~\ref{fg:chi}(a) replotted with $M/H$ on a linear scale.}
\end{figure}

Presenting the $M/H$ data on a logarithmic scale exaggerates the weak response at $T>T_{c2}$.  A direct comparison with other studies is made easier by viewing the data on a linear scale, as in Fig.~\ref{fg:chilin}.   Note that the visibility of the initial diamagnetism below $T_{c1}$ is sensitive to the orientation of the field with respect to the crystal axes.

\begin{figure}[b]
 \centering
    \includegraphics[width=0.95\columnwidth]{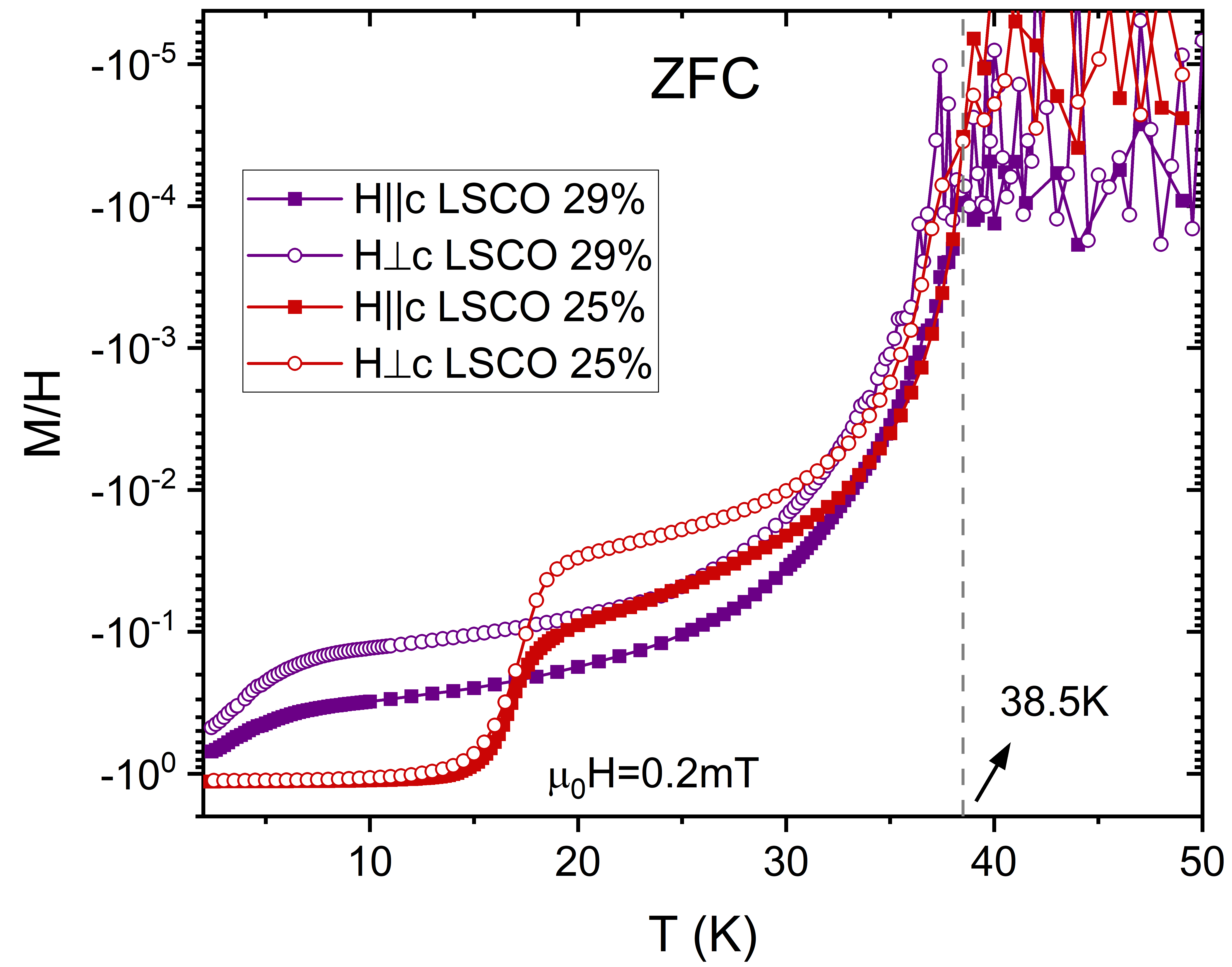}
    \caption{\label{fg:chi29}  Volume susceptibility measured after zero-field cooling (ZFC) as a function of temperature in a magnetic field of 0.2~mT applied parallel to the $c$ axis (filled circles) or perpendicular to the $c$ axis (open squares) for LSCO $x=0.29$ (purple symbols) and $x=0.25$ (red symbols).  (A constant value of $1\times10^{-5}$ has been subtracted from $M/H$ to accommodate plotting on a logarithmic scale, and magnitude has been corrected for the demagnetization factor.) }
\end{figure}

\begin{figure}[t]
 \centering
    \includegraphics[width=0.9\columnwidth]{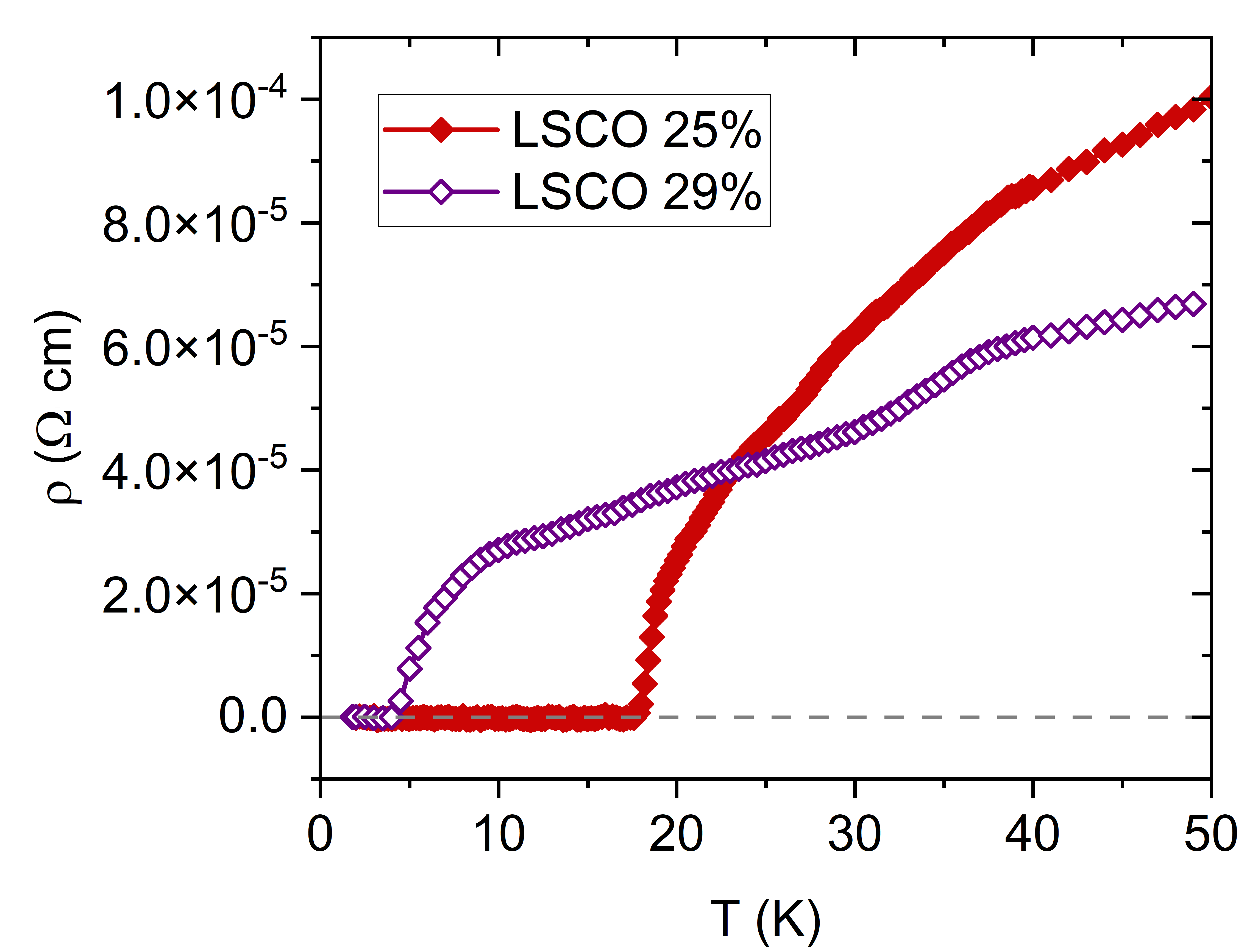}
    \caption{\label{fg:rho29} In-plane resistivity in zero magnetic field for LSCO $x=0.29$ (purple open diamonds) compared with $x=0.25$ (red filled diamonds).}
\end{figure}

The real surprise comes from comparing with the in-plane resistivity data plotted in Fig.~\ref{fg:chi}(c): there is a slight change of slope at $T_{c1}$, but the resistivity only approaches zero at $T_{c2}$.   This is unusual behavior, both for cuprates and for superconductors in general.  Of course, to detect it, one must measure both the resistivity and susceptibility, which is not always done.  It is not uncommon to have samples that exhibit filamentary superconductivity \cite{naga91}; however, in such cases, zero resistivity may be observed at a higher temperature than the onset of bulk magnetic shielding.  An example is a study of La$_2$CuO$_{4+\delta}$, in which the superconducting phase associated with segregation of the excess oxygen was estimated to be $\sim30$\%\ of the volume; the resistivity appeared to drop to zero at $\sim32$~K \cite{koga95}, while the diamagnetism only showed an onset below 31~K \cite{xion96}.  Here we observe the opposite situation.  In the initial phase, a small but significant volume fraction exhibits diamagnetism, but there is no continuous path across the sample that can carry current with zero resistance.

{\newr
These results motivated us to take a careful look at our $x=0.29$ crystal.
Figure~\ref{fg:chi29} compares zero-field-cooled measurements of the volume susceptibility for \lsco\ $x=0.29$ (purple symbols) with the results for $x=0.25$ (red symbols).  Clearly, both samples show an onset of diamagnetism near 38~K; however, for $x=0.29$ the development of a bulk-like response only occurs below 5~K, and full-volume shielding is not achieved at the lowest measurement temperature (1.8~K).

Figure~\ref{fg:rho29} compares the temperature dependence of the in-plane resistivity of the two samples through the superconducting transitions.  The $x=0.29$ crystal starts with a lower normal-state resistivity and then exhibits a slight, gradual drop on cooling below 38~K.  It continues to decrease gradually, going (nominally) to zero at 4~K.  In contrast to the case of $x=0.25$, the lack of full-volume shielding for $x=0.29$ suggests that the zero resistivity is due to filamentary superconductivity.
}

Could our results be consistent with phase-separated grains of superconductor at one value of $p$ embedded in a metallic phase characterized by a significantly larger $p$?  {\newr Our diffraction and STEM characterizations already appear to rule this out; nevertheless, to further} evaluate this possibility, it is useful to compare with an experimental study of an array of Nb islands deposited on a Au film \cite{eley12}.  That study showed a well-defined drop in the resistance measured across the film at a temperature corresponding to superconducting coherence within the isolated Nb islands, with a second drop and approach to zero resistance as coherent Josephson coupling developed through the Au film.  In our case, the in-plane resistivity in Fig.~\ref{fg:chi}(c) shows a gradual and continuous evolution below $T_{c1}$ that is distinct from the example of a mesoscopic array of superconducting islands coupled through a metal presented in \cite{eley12}.

The measured behavior is consistent with the prediction of Spivak, Oreto, and Kivelson \cite{spiv08}.  We observe {\newr a weak} onset of superconductivity in isolated regions starting at a temperature comparable to optimal doping, suggesting that these regions have a lower hole concentration than the average.  The continuous evolution of the resistivity suggests that more regions begin to participate as $T$ decreases, as one would expect for a random distribution of local environments. Zero resistivity is only achieved at a temperature well below the onset of diamagnetism, once superconductivity is induced in the surrounding regions.

\begin{figure*}[t]
 \centering
    \includegraphics[width=2.0\columnwidth]{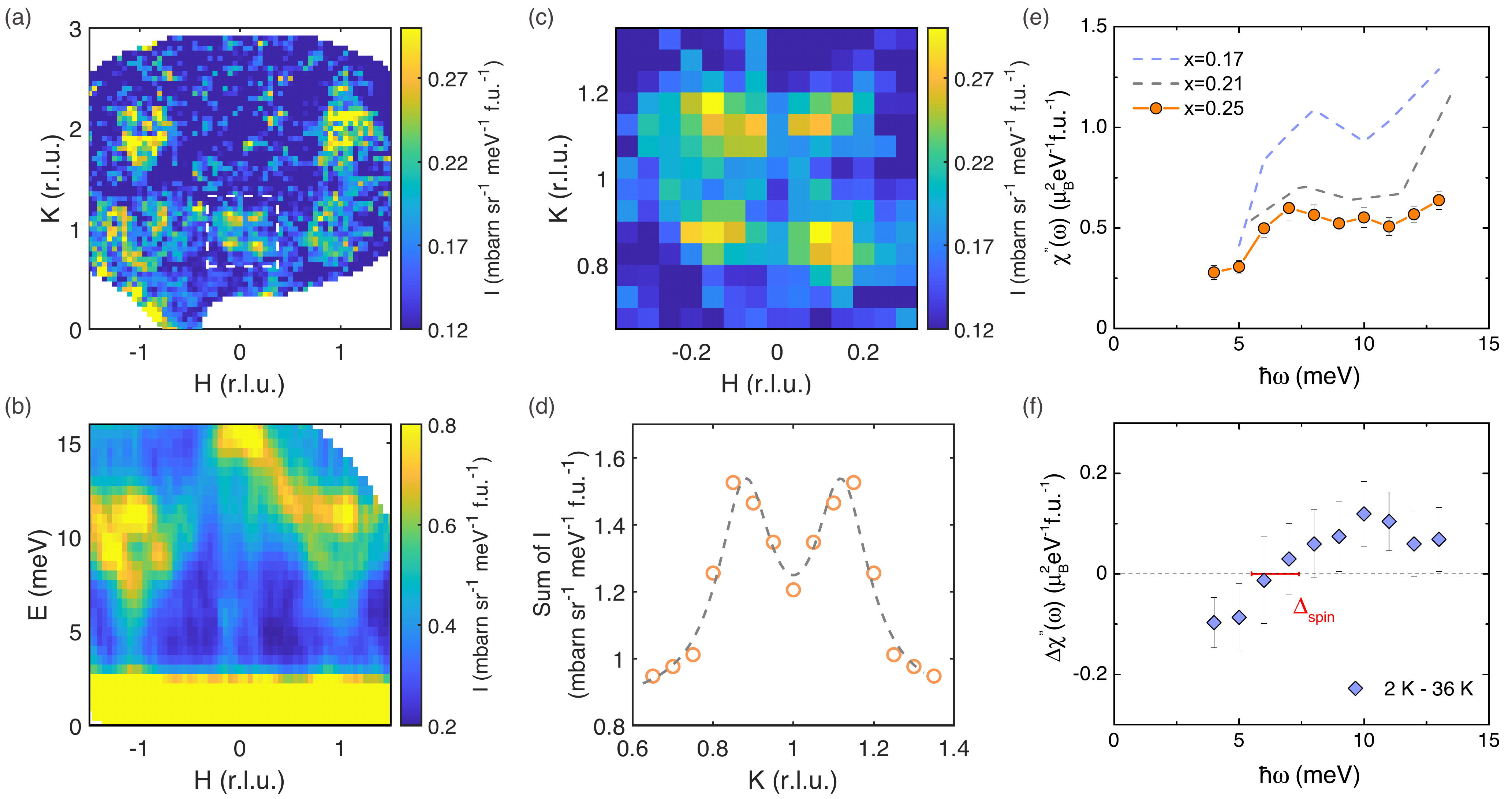}
    \caption{\label{fg:ns} Inelastic neutron scattering measurements on LSCO $x=0.25$.  (a) Intensity within the energy window 6 to 10 meV (and within $L=\pm0.5$) measured at 2 K, after correcting for phonons measured at 300~K, as described in the text.  Incommensurate magnetic scattering can be seen about the antiferromagnetic wave vectors $(0,1,0)$, $(1,2,0)$ and $(-1,2,0)$.  (b) Dispersion of excitations in a slice along ${\bf Q}=(H,1,0)$ (integrated over $\Delta K=\pm0.2$, $\Delta L=\pm0.5$). Besides the magnetic excitations near $H=0$, there are soft phonons at $(\pm1,1,0)$ associated with octahedral tilts. (c) Expansion of the region indicated by the white dashed line in (a).  (d) Cut through (c) for $H = -0.2$ to 0, after symmetrizing about $K=1$.  Dashed line is a fit to a symmetric pair of Lorentzians with centers at $1\pm0.12$ and half-width $\kappa=0.083$~rlu. (e) {\bf Q}-integrated $\chi''(\omega)$ for the present sample (filled circles) compared to results for $x=0.17$ and 0.21 (dashed lines) from \cite{li18}.  (f) Difference in $\chi''(\omega)$ at 2 K relative to 36~K.  The zero-crossing estimate indicated by the red bar corresponds to $\Delta_{\rm spin}=6.5\pm1$~meV.}
\end{figure*}

\section{Magnetic excitations}

The purpose of the inelastic scattering measurements was to characterize the antiferromagnetic spin correlations, but, given the weakness of their scattering, we also have to take account of phonons.  Figure~\ref{fg:ns}(b) shows examples of both signals in a slice of scattering along ${\bf Q}=(H,1,0)$.  The magnetic scattering appears as incommensurate features that extend vertically about the antiferromagnetic wave vector $(0,1,0)$.  Out near the $(-1,1,0)$ and $(1,1,0)$ points, we see evidence of soft phonons associated with rotational fluctuations of the CuO$_6$ octahedra around the Cu-O bonds.  These phonons are in addition to the tilt fluctuations around the diagonal axis of the octahedra that are expected at $(0,1,L)$.  In the low-temperature orthorhombic phase observed for $x\lesssim0.21$ (see App.~A), there are superlattice peaks at $(0,1,L)$ for $L$ even but not equal to 0 \cite{birg87}; in our $x=0.25$ sample with disordered tilts, the fluctuations can spread to lower $L$.  Both of the tilt modes have now been associated with the orthorhombic phase of LSCO \cite{jaco15,sapk21}.  We can correct for much of the phonon response by measuring at 295~K, correcting for the detailed balance factor, and subtracting from the 2 K data.

Figure~\ref{fg:ns}(a) shows a constant-energy slice (integrated from 6 to 10 meV) at 2~K after subtracting the phonon contribution.  This does a good job near $(0,1,0)$, but an imperfect job at $(\pm1,1,0)$ where the soft phonons have their own temperature dependence.  The region around $(0,1,0)$ is shown on a larger scale in Fig.~\ref{fg:ns}(c).  A fit to a symmetrized line cut in Fig.~\ref{fg:ns}(d) shows that the incommensurate peaks appear at $K=1\pm0.12$, consistent with previous work by Wakimoto {\it et al.}~\cite{waki04}.  

To evaluate the strength of the magnetic signal, we plot in Fig.~\ref{fg:ns}(e) the {\bf Q}-integrated imaginary part of the dynamical susceptibility, $\chi''(\omega)$, obtained after correcting for phonon background.  As one can see, this signal is substantially reduced compared to our results for $x=0.17$ and 0.21 \cite{li18}, though it has a similar variation with excitation energy $\hbar\omega$.  

To check the connection with superconductivity, the difference in $\chi''(\omega)$ between 2~K and 36~K is shown in Fig.~\ref{fg:ns}(f) \footnote{Note that we did the neutron experiment before the detailed magnetization study that revealed the initial onset of superconductivity at $T_{c1}$; otherwise we would have done a neutron measurement at 40~K instead of 36~K.  Nevertheless, the diamagnetism at 36~K is still tiny, so that the measurement at this temperature should be very close to the normal-state behavior.}.  It is apparent that magnetic weight is shifted from low energy to high energy.  As in our previous analysis \cite{li18}, we identify the spin gap as the energy at which the change in weight goes through zero, corresponding to $\Delta_{\rm spin}=6.5\pm1$~meV.  As we will discuss, this is somewhat lower than near optimal doping, but large relative to the bulk superconducting transition temperature $T_{c2}$.  (Keep in mind that neutron scattering averages over the entire sample, and we expect a distribution of local environments, only a small fraction of which will contribute to the onset of diamagnetism at $T_{c1}$.) 

\begin{figure*}[t]
 \centering
    \includegraphics[width=1.7\columnwidth]{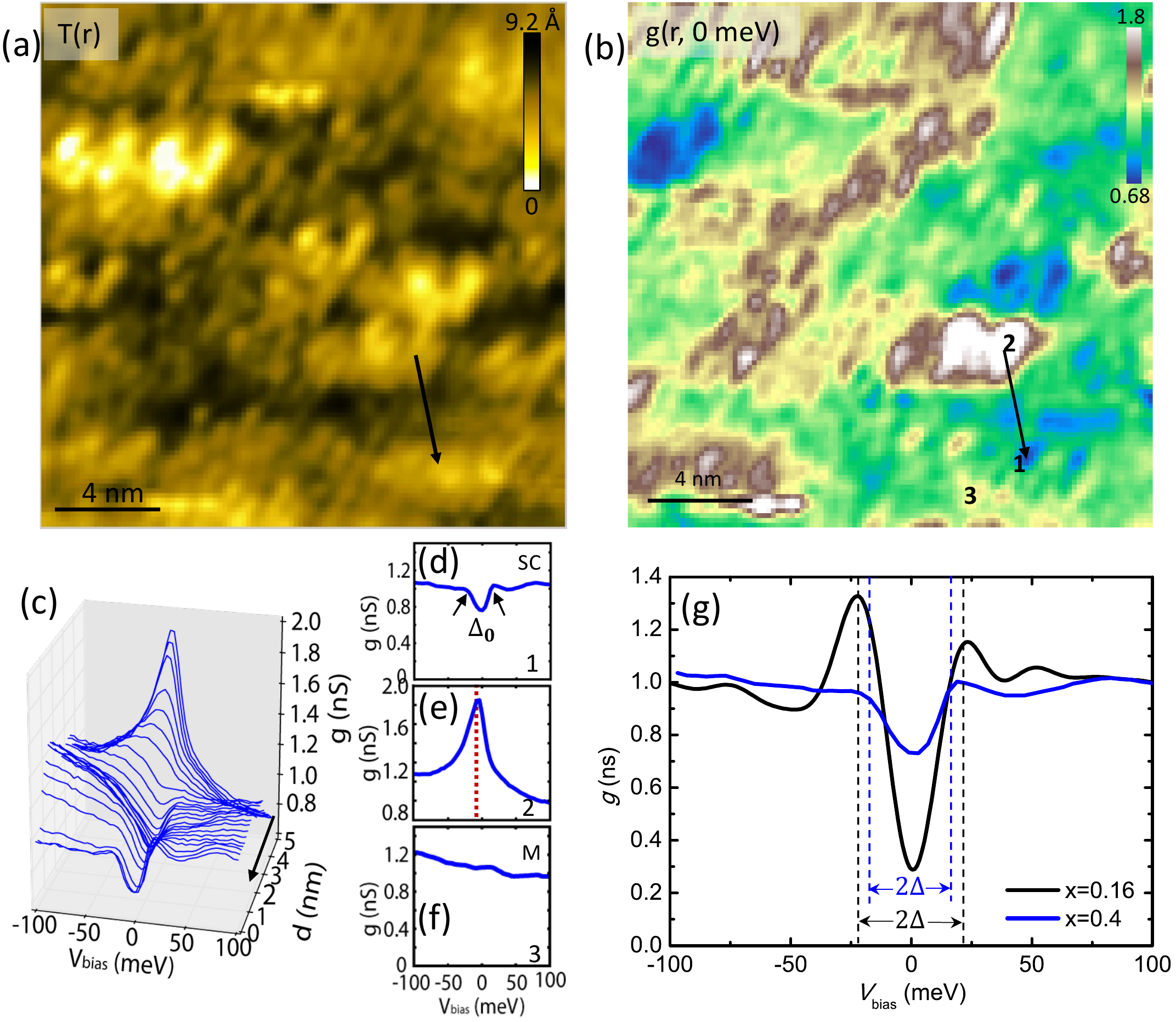}
    \caption{\label{fg:stm}  (a) Topograph, $T(r)$, of a small region of the LSCO film with $p\sim0.35$ ($x=0.4$).  (b) Local density of states at $V_{\rm bias}=0$~mV, $g(r,V_{\rm bias}=0~{\rm mV})$, for the same field of view as in (a).  (c) Differential conductance curves, $g(r,V_{\rm bias})$, measured along the black arrow in (a) and (b).  Select differential conductance curves{\newr , each spatially-averaged over a local region, representative of} the numbered points in (b), shown in (d), (e), (f). {\newr (g) Comparison of  the curve in (d) with the spatially-averaged differential conductance for a superconducting LSCO film with $x=0.16$ \cite{LSCO_STM}.  The dashed lines represent the median energies of coherence peaks from the corresponding spatial distributions.  To enhance the comparison, both curves have been normalized as described in App.~C.  }  }
\end{figure*}

\section{SI-STM on LSCO with $p\sim0.35$}

Turning to the metallic LSCO film with $p\sim0.35$, a typical topograph obtained by SI-STM is shown in Fig.~\ref{fg:stm}(a), covering a 20~nm $\times$ 20~nm field-of-view (FOV).  The surface shows evidence of irregular steps, with local atomic positions distorted from the pattern expected from the bulk structure, possibly due to oxygen sub-stoichiometry. We note that irregular surfaces have also been seen in previous STM studies of LSCO \cite{kato08}, especially on thin films \cite{yuli10}.  Figure~\ref{fg:stm}(b) shows a measure of the local density of states given by the differential conductance at zero bias, $g(r,V_{\rm bias}=0~{\rm mV})$ [obtained simultaneously with the $T(r)$ in Fig.~\ref{fg:stm}(a)].  While it looks rather heterogeneous, the zero-bias conductance is finite everywhere across the FOV, consistent with a metallic system.  For further insight, the differential conductance curves $g(r,V_{\rm bias})$ measured along the black arrow in Fig.~\ref{fg:stm}(b) are plotted in Fig.~\ref{fg:stm}(c), with a few characteristic curves, measured at the numbered points, presented in Fig.~\ref{fg:stm}(d)-(f).  We see that there are points where there is (1) a dip centered at zero bias, (2) a peak at zero bias, and (3) a linear variation across zero bias.  Case (2), with a peak at zero bias, is suggestive of the van Hove singularity that one expects to see at the Lifshitz transition, as the hole-like Fermi surface at smaller $p$ crosses over to an electron-like Fermi surface, which ARPES studies indicate to occur in an average sense at $p \approx 0.2$ \cite{yosh06,park13,hori18,miao21}.  Case (3), with a lack of distinctive features, is compatible with expectations for the metallic phase at $p\sim0.35$ and beyond.  

{\newr
To provide context for evaluating case (1), we compare the differential conductance curve of Fig.~\ref{fg:stm}(d) with a normalized result for a superconducting film with $x=0.16$ in Fig.~\ref{fg:stm}(g); the latter film, with a thickness of 10 unit cells, was measured at 11~K, well below its $T_c$ of 32~K.  In both cases, the distribution of ``coherence'' peak energies has been evaluated, and the dashed lines indicate the median $\Delta$ for each case \cite{LSCO_STM}; these correspond to 18.8 meV (15.7 meV) for $x=0.16$ ($p=0.35$).  These energies are somewhat larger than the values reported in an earlier SI-STM study on LSCO with $x\le0.21$ \cite{kato08}, but are comparable with those measured by ARPES in the antinodal region \cite{yosh09,zhon22,kusp22}.  The coherent superconducting gap, the scale below which spatially-uniform superconducting order occurs \cite{tran21a}, is smaller \cite{musc10,suga13}, limited by the spin gap \cite{li18}, as we will discuss in the next section.

An important difference between these samples is the magnitude of the zero-bias conductance.  The $x=0.16$ sample exhibits a deep minimum at zero bias, as is commonly observed in STM studies on superconducting cuprates \cite{kato08,lang02}.  The much weaker dip of the $p\sim0.35$ sample, which occurs only in limited spatial regions, is reminiscent of observations in \bscco\ (Bi2212) slightly above $T_c$, where local differential conductance can show a reduced zero-bias dip but little reduction in the energies of the ``coherence'' peaks \cite{gome07}.  The evolution of such tunneling spectra across the overdoped regime in (Bi,Pb)$_2$Sr$_2$CuO$_{6+\delta}$ (Bi2201) has been reported recently by Tromp {\it et al.} \cite{trom22}.  They observe a gradual crossover from uniform superconductivity towards superconducting puddles, with the pairing gap filling in (zero-bias conductance rising), rather than the gap energy decreasing, with doping.  Their results are consistent with and complementary to the story presented here.}

\section{Discussion}

We have seen that measurements of magnetic susceptibility and resistivity in LSCO $x=0.25$  provide evidence for isolated regions that develop local superconducting order at a temperature comparable to that of $T_c$ at optimal doping, whereas bulk superconductivity is only established at a much lower $T_{c2}$ that is consistent with previous reports for bulk superconducting order.    
As mentioned in Sec.~I, early work on bulk samples revealed related behavior in magnetization measurements on polycrystalline samples, especially that of Torrance {\it et al.}\ \cite{torr88} and Takagi {\it et al.}\ \cite{taka92a}.   As mentioned in Sec.~II.A, {\newr powder diffraction and STEM characterizations} found no evidence for macroscopic variation in composition.  The observation of similar behavior in single- and poly-crystalline samples with differing synthesis methods provides evidence that it is intrinsic.

The tunneling conductance measurements on the metallic LSCO film ($p\sim0.35$) suggest that, even beyond the superconductor-to-metal transition, regions survive with a gap suggestive of local pairing correlations.  To put these results in context, previous SI-STM studies on LSCO cleaved crystals with $0.06\le x\le0.21$ \cite{kato08} found evidence for a zero-bias-centered gap at all points on the surface and relative homogeneity on the scale of $\pm5$~meV, but considerable disorder on the scale of the coherence peaks (6-20~meV).  At $x=0.21$, the differential tunneling conductance at zero bias has become finite everywhere, suggestive of a partial filling-in of the gap \cite{kato08}.  The variation from that behavior to our observations in Fig.~\ref{fg:stm} due to doping have some parallels with the impact of temperature on the spatial gap distribution detected by SI-STM on Bi2212 \cite{pasu08,park10} {\newr and with a recent study of the doping dependence of gaps in Bi2201 \cite{trom22}.}

\begin{figure}[b]
 \centering
    \includegraphics[width=\columnwidth]{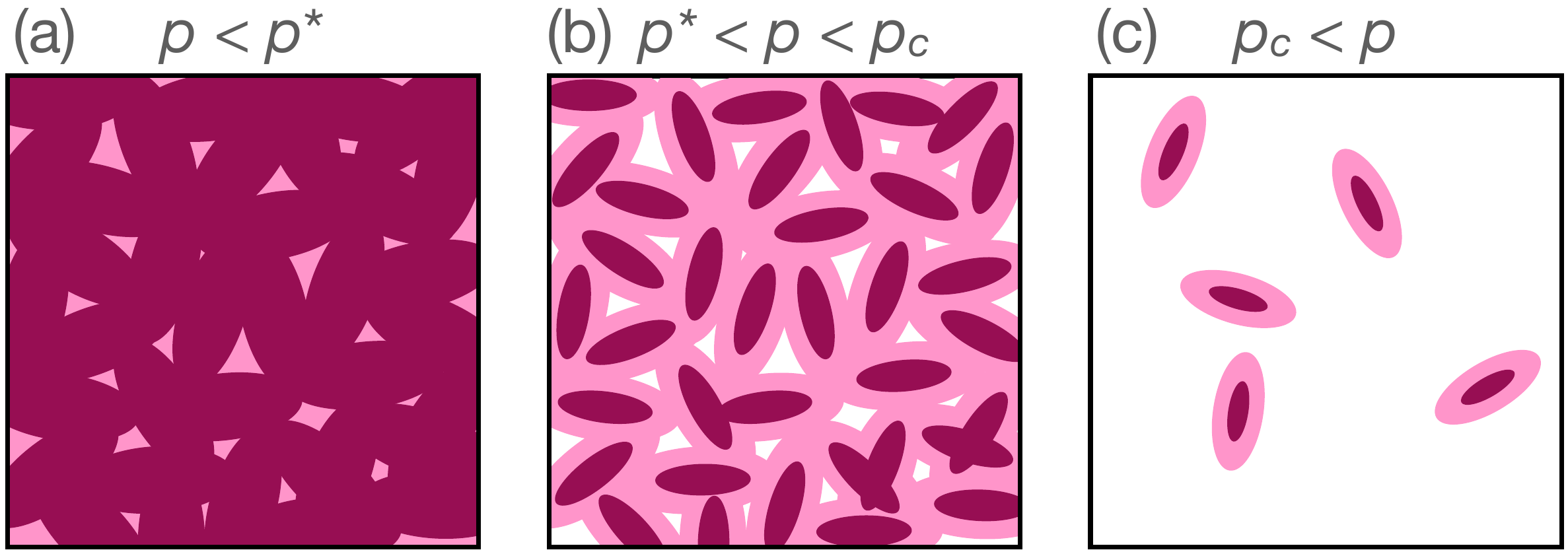}
    \caption{\label{fg:cart}  Cartoon of spatial distribution of pair density within a thin slice of bulk LSCO for (a) $p<p^\ast$, (b) $p^\ast<p<p_c$, and (c) $p_c < p$.  Dark (plum) indicates the highest pair density; light (pink) indicates proximity-induced pair density; white indicates regions where no pairs exist.  }
\end{figure}

In order to compare with the theoretical analyses of Spivak {\it et al.} \cite{spiv08} and Li {\it et al.} \cite{li21c}, we note that the latter paper emphasizes that the combination of disorder and $d$-wave pairing leads to self-organized granularity.  We present a caricature of the evolution of such granularity with doping in Fig.~\ref{fg:cart}, where the depth of color is proportional to pair density.  For $p$ a bit below $p^\ast$, the strongly-correlated grains overlap to form a continuous network with a large pair density.  As $p$ exceeds $p^\ast$, the strongly-correlated grains no longer overlap on average.  The superconducting coherence depends on the proximity-induced superconductivity in the surrounding regions; these have a lower local pair density, resulting in a reduced superconducting phase stiffness. Pockets that are further from the grains will have unpaired carriers.  For $p$ beyond $p_c$, the grains with pairing are too dilute to enable bulk superconductivity.

The neutron scattering measurements demonstrate that the spectrum of magnetic excitations at low energies has a comparable form to that at lower doping, but with reduced magnitude, consistent with previous work \cite{waki04}.  To be more quantitative, we compare in Table I the integral of $\chi''(\omega)$ between 5 and 13 meV, normalized to the result at $x=0.17$.  We see that this measure of the spin correlations is reduced by a factor of 2 for $x=0.25$ compared to $x=0.17$.  Following this trend, previous studies have found negligible low-energy magnetic weight for $x=0.30$ \cite{waki07b,ikeu22}.  

Here we should note that resonant inelastic X-ray scattering (RIXS) measurements on LSCO films have provided evidence for high-energy ($> 100$~meV) magnetic excitations that do not decrease substantially with overdoping, even into the non-superconducting regime \cite{dean13}.  In interpreting these results, it is important to note that the wave vector range probed is closer to $Q = 0$ than to ${\bf Q}_{\rm AF}$.  Determinantal quantum Monte Carlo calculations on the Hubbard model with $U/t=6.5$ are consistent with both neutron and RIXS results: the intensity of the small $Q$ excitations is relatively insensitive to doping while the low-energy excitations close to ${\bf Q}_{\rm AF}$ are very sensitive and disappear with overdoping \cite{huan17b}.  (RIXS measurements with higher energy resolution do show a significant change and anisotropy to the damping of the magnetic excitations with doping \cite{meye17,ivas17,roba19}.)

\begin{table}[t]
\caption{\label{tab}%
Comparison of various quantities measured on LSCO crystals as a function of $x$.  Here, $\int\chi''$ indicates the integral of $\chi''(\omega)$ between 5 and 13~meV, normalized to the result for $x=0.17$.  The uncertainties in $\Delta_{\rm spin}$ and $\int\chi''$ are on the order of 10\%. }
\begin{ruledtabular}
\begin{tabular}{ccccc}
$x$ & $T_c$ & $\Delta_{\rm spin}$ & $2\Delta_{\rm spin}/k_{\rm B}T_c$ & $\int\chi''$ \\
 & (K) & (meV) & & (normalized) \\
\hline
0.17 & 37 & 9.0 & 5.7 & 1.0 \\
0.21 & 30 & 9.0 & 7.0 & 0.7 \\
0.25 & 18 & 6.5 & 8.4 & 0.5 \\
\end{tabular}
\end{ruledtabular}
\end{table}

Applying Occam's razor, we associate the magnetic excitations, which have a similar incommensurability as at $x\sim0.12$ \cite{birg06}, with the correlated regions hosting the isolated superconducting bubbles that appear above $T_{c2}$.  While the decreasing magnetic response is consistent with such an association, the neutron measurements average over the sample volume so that we cannot directly distinguish between uniform and inhomogeneous distributions.  A more relevant parameter is the incommensurate spin gap $\Delta_{\rm spin}$.  In previous work \cite{li18}, it was found that $\Delta_{\rm spin}$ is an upper limit for the coherent superconducting gap, where the latter scales with $T_c$ \cite{deve07,guya08,munn11}.   The coherent gap is a scale below which SI-STM sees spatial homogeneity \cite{tran21a}; for under- to optimally-doped cuprates, that scale can be considerably smaller than that of the local ``coherence'' peaks.   In Table I, we compare the ratio $2\Delta_{\rm spin}/k_{\rm B}T_c$ for several $x$.    If the spin gap limits the coherent superconducting gap \cite{li18,tran21a}, and if $k_{\rm B}T_c$ is proportional to the latter, then we might expect this ratio to remain constant if spin correlations are present in all regions of each sample. Contrary to that, the measured ratio grows with $x$, as the magnetic weight decreases.   Hence, it is plausible to conclude that the superconducting bubbles are correlated patches, and that bulk superconductivity only develops once a coupling between these regions is established through the more conventional metallic areas, associated with larger $p$.

One might argue that, if the correlated regions drive the onset of superconductivity at $T_{c1}$ in our $x=0.25$ sample, then $\Delta_{\rm spin}$ should not decrease from its value at lower doping.  On the other hand, the superconducting onset at 38~K is weak, the typical effective grain size may be decreasing, and the neutron scattering measurements average over the distribution of such regions.  We believe that the comparison of $\Delta_{\rm spin}$ with the bulk $T_{c2}$ is the appropriate one.

The resulting ``plum-pudding'' model of overdoped cuprates has some overlap with the idea of phase separation \cite{wen00,tana05,tana07,wang07,uemu01}, although our perspective is somewhat different.  There is increasing experimental \cite{tran21a} and theoretical \cite{jian22,mai22,wiet22} evidence that charge stripe correlations, defined by antiferromagnetic spin stripe correlations, are relevant to pairing and superconductivity in cuprates.  All doped holes effectively go into charge stripes at low doping, but the stripe density saturates at $p\sim1/8$, so that holes added beyond that point must go into more uniformly-doped regions \cite{birg06}.  As the uniformly-doped regions (``pudding'') grow, the stripe-correlated regions (``plums'') become isolated.

The interpretation of low-energy incommensurate spin excitations in terms of spin stripes is most commonly applied to LSCO and isostructural cuprates \cite{birg06}; nevertheless, a similar evolution of spin correlations is common to other cuprate superconductors, as demonstrated by inelastic neutron scattering studies \cite{enok13,chan16c,li18}.  The character and evolution of high-energy antiferromagnetic excitations is also common among cuprates \cite{stoc10,fuji12a}.  The correlation between antiferromagnetic fluctuations  and superconductivity is implied by the work of Pelc {\it et al.} \cite{pelc19}, who model the superfluid density, especially in overdoped cuprates, as being proportional to the density of localized holes, where a localized hole corresponds to a half-filled Cu $3d_{x^2-y^2}$ orbital and its associated magnetic moment.

\begin{figure}[t]
 \centering
    \includegraphics[width=\columnwidth]{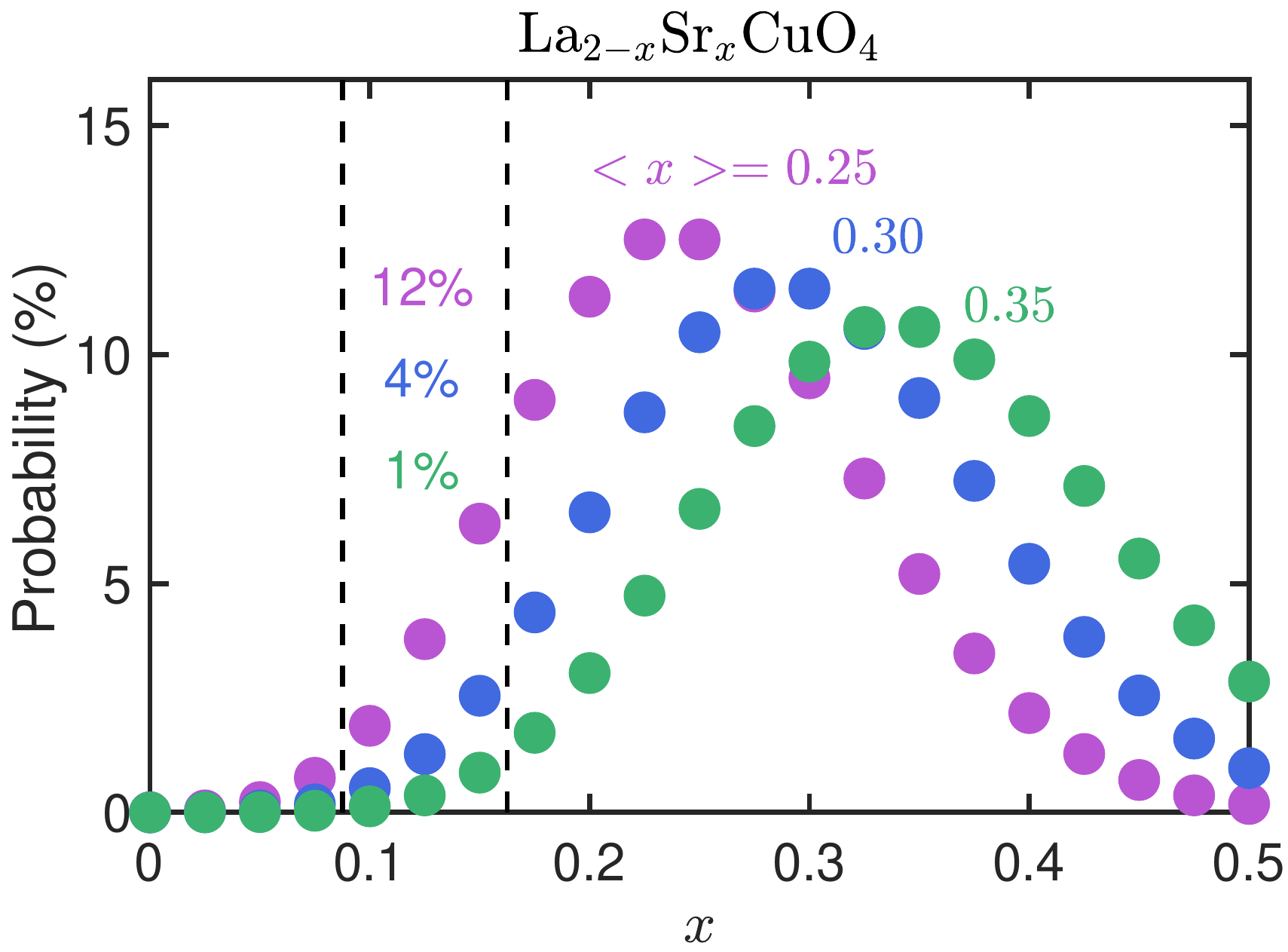}
    \caption{\label{fg:psc}  \newr
    Probability that a region the size of a superconducting coherence volume contains Sr dopants corresponding to a local doping level $x$.  Results are shown for $\langle x\rangle=0.25$ (violet), 0.30 (blue), and 0.35 (green).  The probability that $0.10\le x\le 0.15$ is 12\%, 4\%, and 1\%, respectively.}
\end{figure}

{\newr
To provide further support for this picture, we consider the probability that the coherence volume for a superconducting pair contains a suitably underdoped composition.  The superconducting coherence length is typically associated with the radius of a vortex core; STM studies on as-grown \bscco\ indicate that $\xi_{\rm SC}\sim2.5a_{\rm t}$, where $a_{\rm t}\approx3.8$~\AA\ \cite{hoff02}. (Note that this is also the length scale on which the energies of ``coherence'' peaks are observed to vary in LSCO \cite{kato05}, as well as \bscco\ \cite{howa01,lang02,gome07}.) The coherence length along the $c$-axis tends to be quite short.  To make a reasonable estimate of the hole concentration within a volume based on local Sr density, we need to make some allowance for the spatial range of Coulomb interactions.  Hence, we will take the size in this direction to be equal to the $c$ lattice parameter, so that two layers are included.  The coherence volume is then $\pi\xi_{\rm SC}^2c \approx 20a_{\rm t}^2c$, which contains 40 formula units, including 80 La sites.  The average number of Sr dopants per coherence volume is 10, 12, and 14, for $\langle x\rangle=0.25$, 0.30, and 0.35, respectively.  The corresponding Poisson distributions are plotted in Fig.~\ref{fg:psc}.  If we take the underdoped strongly-correlated, strong-pairing region to be $0.10\le x\le0.15$, then the probability of a coherence volume having such a local composition is 12\%, 4\%, and 1\%\ for $\langle x\rangle = 0.25$, 0.30, and 0.35, respectively.  

We have to note here that, while these numbers are qualitatively consistent with our experimental results, the diamagnetic response of the $x=0.29$ sample in Fig.~\ref{fg:chi29} relative to that for $x=0.25$ at $T\sim 20$--30~K is larger than we would expect.  Of course, plotting $M/H$ on a log scale puts emphasis on this discrepancy, while minimizing the impact of the large reduction in shielding fraction below $T_{c2}$.

An essential feature of our interpretation, consistent with \cite{spiv08}, is that the superconductivity disappears with overdoping because of a loss of regions that have strong pairing interactions.  A consequence is that, while superconductivity may be induced by proximity effect in neighboring regions of higher carrier density, there will also be regions that remain in the normal state.  This is consistent with the change observed in SI-STM differential conductance mapping from underdoped to overdoped regimes.  In underdoped or optimally-doped samples, despite inhomogeneity, the conductance maps are spatially uniform at sufficiently low bias voltages \cite{pan01,howa01,lang02,kato08}, but with overdoping the zero-bias conductance rises and becomes spatially inhomogeneous \cite{kato08,trom22}, with isolated regions of zero-bias dips for $p>p_c$.  This is an important distinction from the model of Pelc {\it et al.} \cite{pelc19}, who discuss disorder on a scale of CuO$_2$ units, but associate the superfluid density in a mean-field fashion with the density of strongly-correlated Cu sites.  They invoke ``intrinsic superconducting gap disorder'' to allow application of percolative scaling to the variation of superfluid density with $T_c$ for overdoped LSCO thin films with $T_c\lesssim12$~K \cite{bozo16}; however, gap disorder is distinct from our model, which asserts that a mixture including regions with no superconducting gap is essential to understanding the disappearance of superconductivity with overdoping.  It is not just disorder, but the loss of pairing interactions that kills superconductivity.
}

Our interpretation of the effective percolative transition of magnetic correlations also provides insight into the evolution of the electronic spectral function measured by ARPES.  The biggest changes with doping occur for the antinodal states near $(\pi,0)$, where ARPES measurements on LSCO show a complete absence of coherent states in the underdoped regime \cite{yosh07}, in correspondence with strong AF correlations.  On doping to $p>p^\ast$, coherent weight appears in the antinodal region, along with the Lifshitz transition \cite{yosh07,chan13,kim18b,miao21}.  In fact, at $x=0.22$ there is sufficient coherent weight that dispersion of antinodal states can be observed perpendicular to the CuO$_2$ planes \cite{hori18,zhon22}.  Nevertheless, observing finite weight at all points on the Fermi surface should not be taken as proof of a uniform Fermi liquid within the CuO$_2$ planes.  Our neutron scattering results on LSCO $x=0.25$ show that patches of strong correlations can coexist within percolating regions of more-conventional, metallic behavior.

Is the picture we have presented for bulk LSCO samples applicable to other cuprates?  One immediate challenge is the case of LSCO thin films, where each composition exhibits only a single sharp transition \cite{lemb10,bozo16}, effectively corresponding to $T_{c2}$.  
Here dimensionality might play a role in two distinct ways.  First, the diamagnetic response in our LSCO $x=0.25$ crystal at $T>T_{c2}$ is three-dimensional (3D), and a detectable response requires that an individual self-organized grain have dimensions comparable to the magnetic penetration depth, which is of order 1000 nm along the $c$-axis \cite{uchi96}.  In contrast, the total thickness of each LSCO film is only $\sim14$~nm.  Second, dimensionality impacts the probability of a self-organized grain having sufficient size to support superconducting phase stiffness.  Here it is useful to consider a percolation model to take account of the inhomogeneity.  Pelc {\it et al.} \cite{pelc18} have applied such a  model to describe the onset of superconductivity with temperature in samples with $p\lesssim p^\ast$, where each spatial patch has a different onset temperature for superconducting order, but all patches are superconducting at low temperature.  In our case, a fraction of patches would remain nonsuperconducting at low temperature.  The site percolation threshold in a simple-cubic lattice (3D) is less than half of that for a square lattice (2D).  Hence, the probability of having a sufficiently large cluster of sites that can sustain local superconducting order should be greater in a bulk crystal than in a nearly two-dimensional film.  Of course, testing for local heterogeneity in LSCO thin films will require new experiments.

Another issue concerns the $T_c \sim n_s$ scaling observed for overdoped LSCO films \cite{bozo16}.   The same behavior was also found in bulk samples of Tl$_2$Ba$_2$CuO$_{6+\delta}$ (Tl2201) by $\mu$SR \cite{nied93,uemu93}.  Can that behavior be explained by a model of effectively granular superconductivity? Phase fluctuations are likely to be important, as they are in underdoped cuprates \cite{emer95a}; however, an analysis of granular superconductivity in a metallic environment has given a slightly different result \cite{imry12}, \footnote{A new analysis of a granular model is able to describe the $T$-dependence of $n_s$ \protect\cite{sant22}.}.   A percolation model \cite{pelc19} involving superconducting and nonsuperconducting patches has been used successfully to model the variation of $T_c$ vs.\ $n_s$ near $p_c$ for the LSCO films \cite{bozo16}.  In the future, it would be interesting to see whether such a model can account for the apparent differences between thin-film and single-crystal samples in the temperature dependence of the diamagnetism and resistivity for $p$ near $p_c$.

Is self-organized granular superconductivity common to the overdoped regime for all cuprate superconductors?  Tl2201 represents an important challenge to this idea \cite{huss13}.   A sharply-defined and complete hole-like Fermi surface was measured by ARPES on an overdoped crystal with $T_c=30$~K \cite{plat05}.  Quantum oscillations have been observed in a crystal with $p=0.30$ and $T_c=10$~K \cite{vign08} and with reduced amplitude at $p=0.27$ ($T_c=26$~K) \cite{bang10}.  For quantum oscillations to be detectable, there must be regions of a sample where a charge carrier can orbit without scattering within a diameter as large as $\sim1000$~\AA\ \cite{bang10}.  At the same time, the measurement does not prove that all regions of the sample are weakly correlated, just that there must be some sufficiently large regions of ``pudding'' that are not disrupted by ``plums''.  Could there be some ``plums''?  Recent measurements with resonant inelastic X-ray scattering found evidence for charge-density modulation in crystals with $p=0.23$ and 0.25 ($T_c=56$ and 45~K, respectively) \cite{tam22}.  In-plane resistivity varies smoothly across the overdoped range \cite{huss13}, so it seems likely that a few ``plums'' are still present at the largest $p$ where superconductivity survives.

Transport measurements may not be an absolute test for sample homogeneity.  For example, remarkable transport results have been reported for magic-angle twisted-bilayer graphene, where an entire phase diagram, including superconductivity, has been mapped simply by varying the gate voltage \cite{cao18}.  Despite the observation of quantum oscillations in resistivity at fairly low magnetic fields, nanoscale mapping of Landau levels with a scanning probe has imaged substantial disorder in the local twist angle \cite{uri20}.   In this context, it is relevant to note that measurements of in-plane thermal conductivity and resistivity in a superconducting Tl2201 crystal with $p$ close to $p_c$ indicate that the normal-state response satisfies the Wiedemann-Franz law in the $T\rightarrow0$ limit, consistent with transport of heat and charge by quasiparticles, while the resistivity exhibits a $T$-linear term that is inconsistent with Fermi-liquid theory \cite{prou02}.

Another system in which overdoped samples have been investigated is \bscco, where a
detailed ARPES study of the variation of the spectral function for antinodal states across the superconductor-to-metal transition has been made \cite{vall20}.  The large self-energy present at lower $p$ gradually disappears as the transition is crossed.  If one assumes that the electronic response of the sample is homogeneous, then the results suggest a gradual reduction in coupling to the excitations responsible for pairing \cite{vall20}.  On the other hand, if one allows for inhomogeneity \cite{spiv08}, this corresponds to the disappearance of patches with strong AF spin correlations and their impact on regions of metallic quasiparticles.  In fact, it should be possible to check for inhomogeneity in the pair density by SI-STM on such highly-doped \bscco\ samples, and this will be an important test for future work.

Finally, it is interesting to note that Berben {\it et al.} \cite{berb22} have recently reported evidence that strange-metal behavior (in the form of in-plane magnetoresistance scaling as $H/T$) occurs throughout the overdoped regime for three different cuprate families, including LSCO.  Furthermore, Yang {\it et al.}\ \cite{yang22} have reported that artificial disorder applied to an optimally-doped \ybco\ thin film led to the observation of strange-metal behavior.  Could it be that a heterogeneous mixture of local electronic environments is key to the anomalous transport?  This is another topic for future investigation.

In conclusion, we have presented evidence that superconductivity in overdoped LSCO crystals is driven by finite-sized strongly-correlated regions with character similar to what dominates at $p<p^\ast$.  Consistent with the analysis of Spivak {\it et al} \cite{spiv08}, bulk superconductivity appears to be induced by proximity effect in the surrounding weakly-correlated metal.  When, with sufficient doping, the strongly-correlated regions become too dilute, superconducting order disappears. Testing the generality of this picture for cuprates will require new experiments.

\section{Acknowledgments}

We thank S. A. Kivelson and M.-H. Julien for valuable comments.
Work at Brookhaven is supported by the Office of Basic Energy Sciences, Materials Sciences and Engineering Division, U.S. Department of Energy (DOE) under Contract No.\ DE-SC0012704.   A portion of this research used resources at the Spallation Neutron Source and the High Flux Isotope Reactor, DOE Office of Science User Facilities operated by Oak Ridge National Laboratory. {\newr Another portion of this research used resources of the Advanced Photon Source, a DOE Office of Science user facility operated by Argonne National Laboratory under Contract No. DE-AC02-06CH11357. }

\appendix

\section{Structural characterization of crystals}

\begin{figure}[t]
 \centering
    \includegraphics[width=0.9\columnwidth]{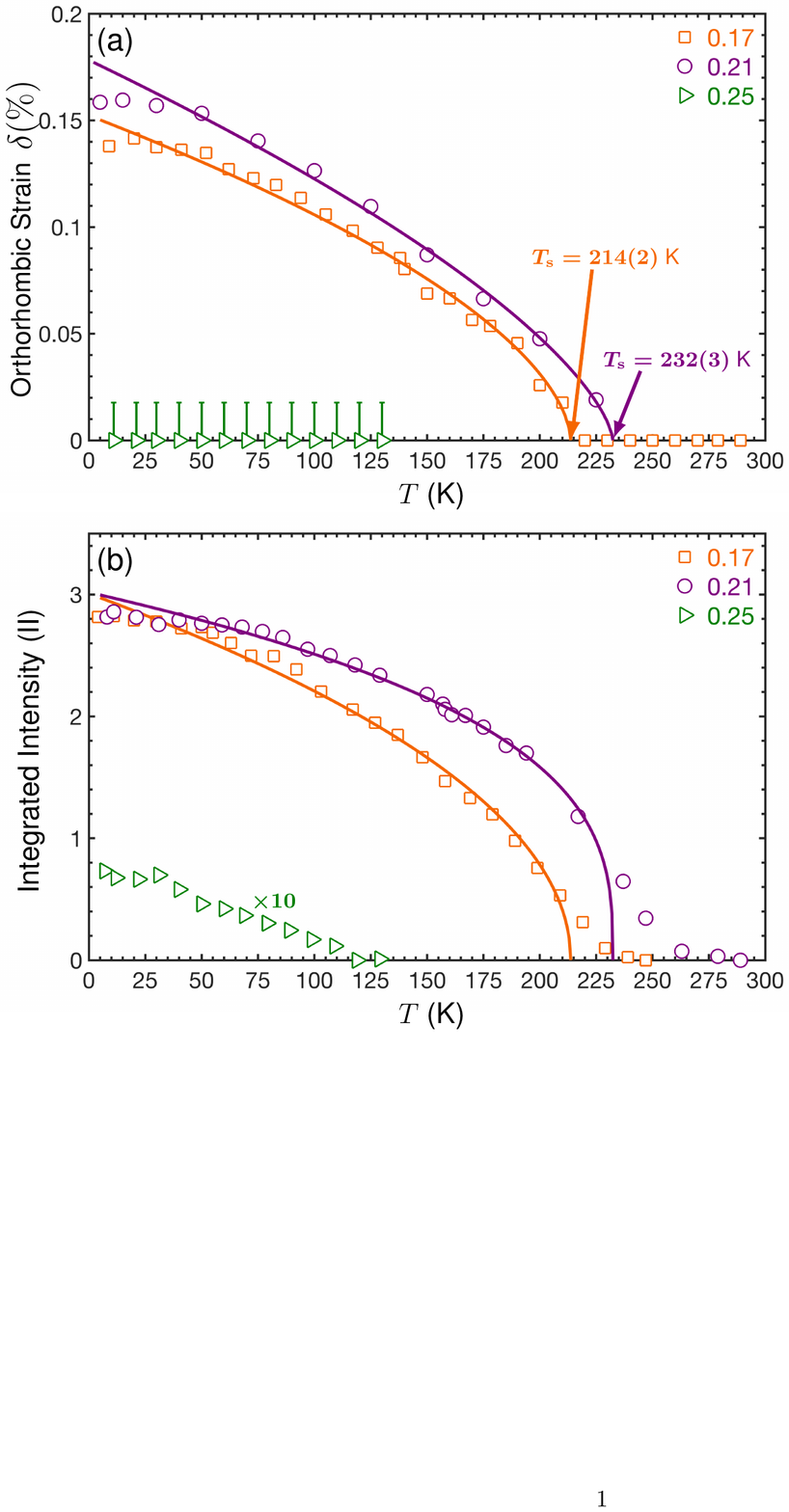}
    \caption{\label{fg:struc} (a) Orthorhombic strain, $\delta=(b-a)/[0.5(b+a)]$, vs.\ temperature for crystals of LSCO $x=0.17$, 0.21, and 0.25, determined from twinned $(2,0,0)/(0,2,0)$ reflections.  Error bars for the $x=0.25$ results indicate the resolution-limited uncertainty.  (b) Integrated intensity of the $(0,3,2)$ orthorhombic superlattice peak normalized to the $(2,0,0)$ integrated intensity for $x=0.21$ and 25.  For $x=0.17$, the integrated $(3,2,1)$ intensity is normalized to the $x=0.21$ result at base temperature.  The fitted curves (lines) are proportional to $(1-T/T_s)^{2\beta}$, with $\beta$ values in the range of 0.16 to 0.33.   }
\end{figure}

An initial measurement of single-crystal neutron diffraction for the $x=0.21$ sample on the Wide-Angle Neutron Diffractometer (HB-2C) at the High Flux Isotope Reactor (HFIR at Oak Ridge National Laboratory) provided clear evidence of orthorhombic superlattice peaks at $T=200$~K, which was a surprise, as discussed below.  This observation motivated us to determine the temperature dependence of the structure of each of the three compositions ($x=0.17$, 0.21, and 0.25) on triple-axis spectrometers at HFIR.  There, each sample was mounted in a closed-cycle He refrigerator, with the $c$ axis approximately vertical.  The $x=0.21$ crystal was studied on HB-1, where twinned $(2,0,0)/(0,2,0)$ reflections were measured with a Si (111) analyzer and horizontal collimations of $48'$-$20'$-$20'$-$30'$.  The intensity of the $(0,3,2)$ superlattice peak was measured with a PG (002) analyzer and $48'$-$40'$-$40'$-$120'$ horizontal collimations; the peak was reached by tilting the crystal with the sample goniometer.  The $x=0.17$ and 0.25 samples were studied on HB-3, with a Si(111) monochromator, PG(002) analyzer, and collimations of $48'$-$60'$-$60'$-$120'$.  All measurements were done with a neutron energy of 13.5~meV. For the $x=0.17$ sample, the $(0,3,2)$ reflection could not be reached, so $(3,2,1)$ was measured instead.  The results for the orthorhombic strain, $\delta=(b-a)/[0.5(b+a)]$, and the relative intensities of the superlattice reflections as a function of temperature are plotted in Fig.~\ref{fg:struc}.

The detectable low-temperature orthorhombic strain and high orthorhombic-tetragonal transition temperature of the $x=0.21$ sample are different from the early powder diffraction work \cite{rada94}, which found the structural transition to reach $T=0$ for $x\approx0.21$.   Of course, the structure is sensitive to oxygen content \cite{torr88} and could be sensitive to grain size, as structural differences between powders and crystals of the related system \lnsco\ have been reported previously \cite{tran99a}.  That the strain and structural-transition temperature are slightly higher for $x=0.21$ than for 0.17 is unexpected, as these quantities are typically observed to decrease monotonically with doping \cite{rada94}.  In contrast to this unusual feature, superconducting transition temperatures of these crystals, as indicated in Fig.~\ref{fg:tc}, are comparable with those from other recent single-crystal studies \cite{wang07}.  Furthermore, angle-resolved photoemission measurements \cite{miao21} demonstrate that the $x=0.17$ and 0.21 crystals are on opposite sides of the Lifshitz transition ($x_c\approx0.2$), where the nominal Fermi surface changes from hole-like to electron-like \cite{yosh06,hori18}.

What is the significance of this difference in transition temperature with doping?  The impact of the substitution of Sr$^{2+}$ ions for La$^{3+}$ is to create local disorder that impacts the long-range ordering of octahedral tilts.  As in the anisotropic random-field Ising model \cite{zach03}, increasing disorder leads to a reduction in the ordering temperature of the orthorhombic phase.  


\begin{figure}[b]
 \centering
    \includegraphics[width=0.6\columnwidth]{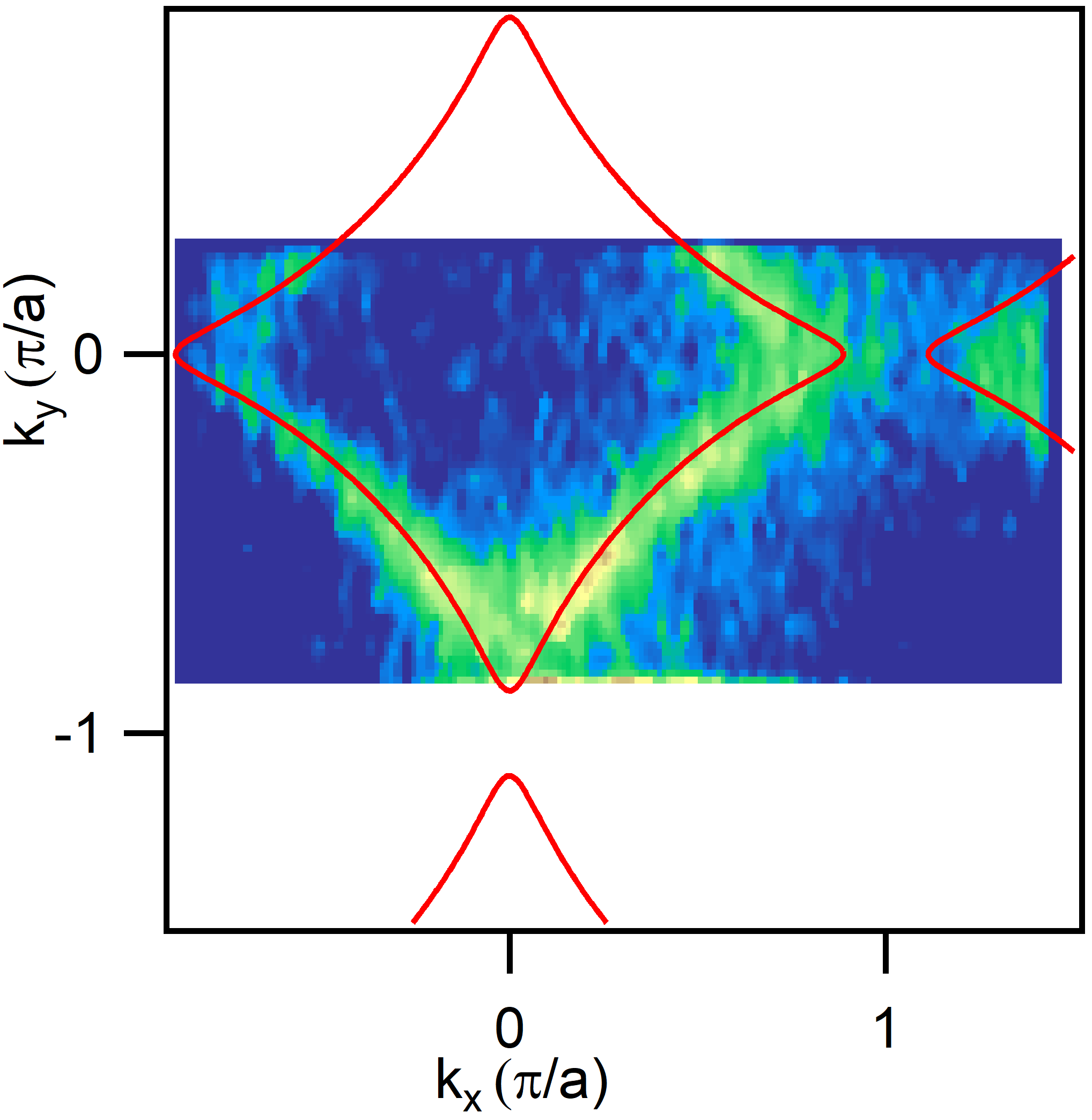}
    \caption{\label{fg:fs}   ARPES intensity map of the LSCO film, integrated over $E_F\pm15$~meV. The solid curve represents a tight-binding fit to the mapped Fermi surface. The doping level is estimated to be $p = 0.35$. }
\end{figure}

\section{ARPES characterization of LSCO film}

After synthesis by molecular beam epitaxy, the LSCO film was transferred {\it in vacuo} to a second chamber for measurements of angle-resolved photoelectron spectroscopy (ARPES).  An ARPES intensity map at the Fermi energy, $E_F$, obtained with a He II photon source (40.8~eV) at room temperature is shown in Fig.~\ref{fg:fs}.  (The reason for measuring at room temperature, rather than low temperature, was to avoid any condensation on the film's surface prior to the transfer to the STM chamber; however, note that the high measurement temperature leads to an effective experimental energy resolution of $\sim95$~meV. )  The solid line is a tight-binding fit to the Fermi surface using the model dispersion of \cite{esch03}.  The average hole concentration can be determined from the area of the Fermi surface \cite{droz18}, resulting in the value $p=0.35$.


{\newr
\section{SI-STM on LSCO $x=0.16$}

We label the measured, spatially-averaged differential conductance for the $x=0.16$ film as $g_{\rm raw}(V)$.  To generate a background for normalization of the data, we make use of the Fermi-Dirac function,
\begin{equation}
  f(V) = {1 \over e^{V/kT} + 1},
\end{equation}
and its derivative with respect to $V$, $f'(V)$.  Taking $T=46$~K, we estimate a nonsuperconducting conductance $g_{\rm bg}$ by convolving the measured conductance with $f'(V)$:
\begin{equation}
  g_{\rm bg} = \int d\omega g_{\rm raw}(\omega) f'(V-\omega).
\end{equation}
To obtain the normalized conductance, $g_{\rm norm}(V)$, that is presented in Fig.~6(g) of the main text, we use:
\begin{equation}
  g_{\rm norm}(V) = g_{\rm raw}(V) / g_{\rm bg}(V).
\end{equation}
The raw and background conductances are shown in Fig.~\ref{fg:stm16}.

\begin{figure}[h]
 \centering
    \includegraphics[width=0.75\columnwidth]{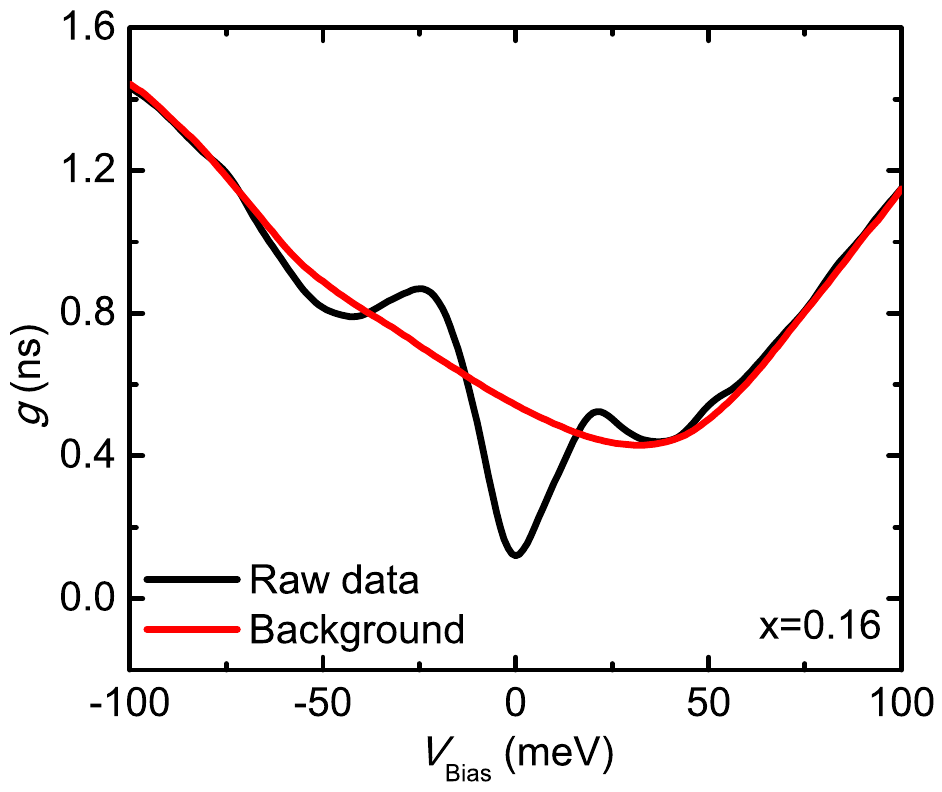}
    \caption{\label{fg:stm16} \newr Black curve shows the measured, spatially-averaged tunneling conductance curve $g_{\rm raw}(V)$ for the LSCO $x=0.16$ thin film.  The red curve is $g_{bg}(V)$ obtained from $g_{\rm raw}(V)$ by convolving with the derivative of the Fermi-Dirac function evaluated with $T = 46$~K, as indicated in Eq.~(2).}
\end{figure}
}
\vfil \null

\bibliography{LNO,theory,neutrons,overdoped}

\end{document}